\documentclass[twocolumn,superscriptaddress,floatfix,preprintnumbers,prd,nofootinbib]{revtex4}
\usepackage[colorlinks=true,breaklinks=true]{hyperref}
\usepackage[utf8]{inputenc}
\hypersetup{urlcolor=Blue,
	    citecolor=Blue,
	    linkcolor=Blue}
\usepackage{mathtools}
\usepackage{amssymb}
\usepackage[usenames,dvipsnames]{xcolor}
\usepackage{orcidlink}

\newcommand{\bx}{{\bf x}}

\newcommand{\bv}{{\bf v}}

\newcommand{\bp}{{\bf p}}

\newcommand{\vc}{v_{\rm c}}

\newcommand{\uR}{u_{\rm R}}
\newcommand{\uI}{u_{\rm I}}

\newcommand{\exclude}[1]{{}}
\long\def\exclude#1{}

\newcommand{\GF}{G_{\rm F}}

\newcommand{\edit}[1]{{\color{black}{#1}}}
\newcommand{\wI}{\omega_{\rm I}}
\newcommand{\wR}{\omega_{\rm R}}

\begin{document}

\title{Fast Flavor Pendulum: Instability Condition}

\author{Damiano F.\ G.\ Fiorillo \orcidlink{0000-0003-4927-9850}} 
\affiliation{Deutsches Elektronen-Synchrotron DESY,
Platanenallee 6, 15738 Zeuthen, Germany}

\author{Manuel Goimil-García \orcidlink{0009-0001-0518-9274}
} 
 \affiliation{Niels Bohr International Academy \& DARK, Niels Bohr Institute,\\University of Copenhagen, Blegdamsvej 17, 2100 Copenhagen, Denmark}

\author{Georg G. Raffelt \orcidlink{0000-0002-0199-9560}}
\affiliation{Max-Planck-Institut f\"ur Physik, Boltzmannstr.~8, 85748 Garching, Germany}


\begin{abstract}
Even in the absence of neutrino masses, a neutrino gas can exhibit a homogeneous flavor instability that leads to a periodic motion known as the fast flavor pendulum. A well-known necessary condition is a crossing of the angular flavor lepton distribution. \edit{In an earlier work, some of us showed that homogeneous flavor instabilities also obey a Nyquist criterion, inspired by plasma physics. This condition, while more restrictive than the angular crossing, is only sufficient if the unstable branch of the dispersion relation is} bounded by critical points that both lie under the light cone (points with subluminal phase velocity). While the lepton-number angle distribution, assumed to be axially symmetric, easily allows one to determine the \edit{real-valued} branch of the dispersion relation and to recognize if instead superluminal critical points exist, this graphical method does not translate into a simple instability condition. We discuss the homogeneous mode in the more general context of \edit{the dispersion relation for} modes with arbitrary wave number and stress that it plays no special role on this continuum, except for its regular but fragile long-term behavior, owed to its many symmetries.
\end{abstract}

\maketitle

\section{Introduction}
\label{sec:intro}

In a dense neutrino gas, neutrino-neutrino refraction spawns flavor waves that can be unstable and then lead to flavor conversion. These unstable collective modes do not require flavor mixing and not even neutrino masses, in which case they are termed fast flavor modes, namely space and time dependent solutions of the underlying quantum kinetic equations. A question of both theoretical and practical interest is the condition for the appearance of such run-away modes of flavor conversion. In the fast flavor case, a crossing of the angular distribution of flavor lepton number carried by neutrinos is a well-known necessary and sufficient condition \cite{Morinaga:2021vmc, Dasgupta:2021gfs, Johns:2024bob, Fiorillo:2024bzm}.

However, one often considers systems that have various degrees of symmetry, notably axially symmetric distributions as well as axially symmetric solutions, in which case Morinaga's theorem \cite{Morinaga:2021vmc, Fiorillo:2024bzm} does not apply because the guaranteed unstable modes break axial symmetry. For a single crossing, an axially symmetric unstable solution is still guaranteed, as well as for an odd number of crossings~\cite{Fiorillo:2024uki}, whereas for an even number of crossings, all axially symmetric solutions could be stable~\cite{Capozzi:2019lso}.

One particularly eye-catching solution, both axially symmetric and homogeneous, is known as the fast flavor pendulum \cite{Johns:2019izj, Padilla-Gay:2021haz, Fiorillo:2023mze,Fiorillo:2023hlk}. This particularly symmetric case is special because it is technically integrable due to its large number of conserved quantities, the Gaudin invariants \cite{Pehlivan:2011hp, Raffelt:2011yb, Fiorillo:2023mze,Fiorillo:2023hlk}, although in practice these symmetries will be spontaneously broken by the coupling to higher-wavenumber modes and by collisions that dissipate flavor coherence \cite{Mangano:2014zda, Padilla-Gay:2022wck,Fiorillo:2023ajs}. Despite its extreme fragility, the fast flavor pendulum remains an intriguing exact solution, the prototype of a fast instability, and as such commands particular attention.

Besides the necessity of an angular crossing, which properties of the angle distribution would guarantee the appearance of this solution? As the general dispersion relation involves various integrals over the angle distribution, there is no simple answer. For a given real wave number $k$ along the symmetry axis, the potentially complex frequency of the corresponding normal mode is $\omega(k)$ and instability requires $\wI(k)={\rm Im}\,\omega(k)>0$. For a single-crossed spectrum, the existence of a $k$-interval with complex $\omega(k)$ is guaranteed, but not that it includes the homogeneous case. 

Actually, for any single-crossed spectrum, in all cases we have explored, there are either two intervals $(k_1,k_2)$ and $(k_3,k_4)$ of unstable modes with \hbox{$k_1<k_2<k_3<k_4$}, or only a single interval $(k_1,k_4)$, as first suggested by the examples of Ref.~\cite{Yi:2019hrp}. In both cases, the limiting points $\omega(k_1)$ and $\omega(k_4)$ lie under the light cone, i.e., they have subluminal phase velocity $|\wR(k_{1,4})/k_{1,4}|<1$, where $\wR(k)={\rm Re}\,\omega(k)$. If the points $k_{2,3}$ exist, the end points $\omega(k_{2,3})$ are superluminal. In other words, if there is a single unstable interval, it begins and ends under the light cone (but may be outside in the intermediate range). If there are two intervals, each of them has a subluminal and a superluminal end point. Two of us have proven these properties in the more general context of inhomogeneous, axially-breaking instabilities \cite{Fiorillo:2024uki}. However, \edit{it remains mathematically unproven whether these properties are truly generic for axially symmetric solutions or~not.}

Previously, using methods of plasma physics, two of us have developed the ``Nyquist criterion'' for the homogeneous mode to be unstable \cite{Fiorillo:2023hlk}, and it was thought to be both necessary and sufficient. However, it was effectively assumed that the endpoints of the unstable branches are under the light cone, which implies that some neutrinos are on resonance with the flavor wave. The possibility of the branches ending above the light cone was missed because in the physics of nonrelativistic plasmas, some of the electrons are always on resonance with the plasma wave. So with hindsight, the Nyquist criterion only applies if
there is a single unstable interval $(k_1,k_4)$, whereas in the case of two intervals, it is still necessary, but no longer sufficient. Therefore, to understand the conditions for the homogeneous mode to become unstable, one cannot avoid a general study of the dispersion relation for all wavenumbers; this serves in itself as a reminder that the homogeneous mode should not really be regarded as special or more instructive---its beauty and simplicity can be deceiving.

Therefore, while clarifying the status of the Nyquist criterion, we use this opportunity to discuss more explicitly the dispersion relation that follows from a single-crossed axially symmetric angle distribution, and what it implies specifically for the homogeneous mode.  In Sec.~\ref{sec:Dispersion} we recap the dispersion relation for an axisymmetric fast flavor system. In Sec.~\ref{sec:Nyquist} we explain the Nyquist criterion, which remains a necessary condition and is more restrictive than simply a crossing of the angle distribution. In Sec.~\ref{sec:Example} we illustrate the typical properties of the dispersion relation with a family of distributions, somewhat parallel to Ref.~\cite{Yi:2019hrp}\edit{, and check whether the Nyquist criterion is fulfilled}. We also prove the general properties of the single-crossed dispersion relation in Sec.~\ref{sec:General} and discuss what they mean for the homogeneous mode. We follow the formal results introduced in Ref.~\cite{Fiorillo:2024uki}, but here we apply them practically to specific angular distributions to show how to actually determine whether a specific wavenumber is unstable or not. We finally conclude in Sec.~\ref{sec:Conclusion}.

\section{Dispersion relation}
\label{sec:Dispersion}

\subsection{Axisymmetric equations of motion}

The quantum-kinetic description starts with the flavor density matrices $\varrho_\bp(\bx,t)$, where the diagonal entries are the usual occupation numbers, whereas the off-diagonal elements $\psi_\bp(\bx,t)$ and complex conjugates encode flavor coherence. In the linear approach, $\psi_\bp(\bx,t)$ is what we call flavor field and for which we seek the dispersion relation. In general, there are three such fields for the coherence between any pair of flavors, but in the linear limit we can study each of them separately---in the linear limit there are no three-flavor effects \cite{Airen:2018nvp}. In general, there are independent density matrices $\overline\varrho_\bp(\bx,t)$ for antineutrinos and corresponding flavor fields.

The fast flavor limit is defined by neglecting neutrino masses, implying that the density matrices for lepton number $\rho_\bp\equiv\varrho_\bp-\overline\varrho_\bp$ obey self-contained equations of \hbox{motion} (EOMs) which descend from the quantum kinetic equation~\cite{Dolgov:1980cq, Rudsky, Sigl:1993ctk, Sirera:1998ia, Yamada:2000za, Vlasenko:2013fja, Volpe:2013uxl, Serreau:2014cfa, Kartavtsev:2015eva, Fiorillo:2024fnl, Fiorillo:2024wej}. After neglecting background matter and collision effects, they are
\begin{equation}
    (\partial_t +\nabla_\bx)\rho_\bp
    =-i\sqrt{2}\GF \int \frac{d^3\bp'}{(2\pi)^3}\bigl[\rho_{\bp'},\rho_\bp\bigr](1-
    \bv\cdot\bv'),
\end{equation}
where $\bv=\bp/|\bp|$ is the velocity vector, a unit vector that represents the direction of motion. The space-time dependence of the $\rho_\bp$ matrices is not explicitly shown. Moreover, the EOMs do not depend on energy so that we can use energy-integrated lepton-number density matrices that depend only on $\bv$ instead of $\bp$. We neglect any collisional term potentially connected with novel branches of collisional instabilities~\cite{Johns:2021qby,Xiong:2022zqz, Liu:2023pjw, Lin:2022dek, Johns:2022yqy, Padilla-Gay:2022wck, Fiorillo:2023ajs}.

Assuming axial symmetry around the $z$-direction, the azimuth angle is integrated out, leaving a dependence only on $v=\cos\theta$. Therefore, we may use integrated density matrices $\rho_v(z,t)$ that depend only on $v$ as well as on $z$ and $t$. Moreover, we normalize them to the total neutrino density so that $\rho_v$ is understood to stand for $(n_\nu+n_{\overline{\nu}})^{-1}\int \rho_\bp \bp^2\,d{|\bp|}\,d\phi/(2\pi)^3$. The simplified EOMs then read
\begin{equation}
\label{eq:qke}
    (\partial_t+v\partial_z)\rho_v
    =-i\mu \int_{-1}^{+1}\!dv' \bigl[\rho_{v'},\rho_v\bigr](1-vv'),
\end{equation}
where $\mu=\sqrt{2}\GF(n_\nu+n_{\overline{\nu}})$ is the effective neutrino-neutrino interaction strength. In the literature it is often defined asymmetrically, using $n_\nu$ only.

\subsection{Linearization}

Next we assume that initially neutrinos are in flavor eigenstates and thus that $\rho_v$ is diagonal aside from small seeds and we also assume that the diagonal elements are perfectly homogeneous. Therefore, on the linear level, the trace of each $\rho_v$ is conserved, allowing us to focus only on the small off-diagonal pieces and we write
\begin{equation}
    \rho_v-\frac{1}{2}
        \begin{pmatrix}1 & 0\\ 0 & 1\end{pmatrix}  {\rm Tr}\,\rho_v
    =\frac{1}{2}\begin{pmatrix}
        G_v & \psi^*_v(z,t)\\
        \psi_v(z,t) &-G_v
    \end{pmatrix},
\end{equation}
where $G_v$ is the angular spectrum that is independent of $z$ and $t$ and $\psi_v(z,t)$ is the space-time dependent flavor field. If one expresses the density matrix in terms of the usual polarization vectors, $G_v=P_v^z$ and $\psi_v=P_v^x+iP_v^y$, following our previous works~\cite{Fiorillo:2024bzm, Fiorillo:2024uki}. In the literature, often the upper-right component $\psi_v^*=P_v^x-iP_v^y$ was taken as the flavor field, explaining certain sign differences.

With these assumptions and notations, we finally find the linear EOMs in the form
\begin{equation}\label{eq:eom}
    (\partial_t+v\partial_z)\psi_v=i\mu
    \bigl[\psi_v(G_0-vG_1)-G_v(\psi_0-v\psi_1)\bigr].
\end{equation}
We use the angular moments $G_n=\int dv\, G_v v^n$ and $\psi_n=\int dv\, \psi_v v^n$, where only the monopole and dipole terms appear in the EOMs.\footnote{Usually we express the dependence on $v$ as a subscript such as $G_v=G(v)$, which typically should not lead to confusion with the moments $G_0$, $G_1$ that do not signify $G(0)$ or $G(1)$. In a few instances we actually need $G(\pm1)$, the value of $G(v)$ at the edges of the interval $-1\leq v\leq +1$, and then we write explicitly $G(\pm1)$.\label{fn:notation}}

\subsection{Normal modes}

One may next seek normal modes that fulfill these EOMs, namely plane-wave solutions of the form $\psi_v\to \psi_v e^{-i\Omega t+iK z}$, where the notation for $\psi_v(z,t)$ and the normal-mode amplitude $\psi_v(\Omega,K)$ is the same. Moreover, we introduce the shifted frequency $\omega=\Omega+\mu G_0$ and wavevector $k=K+\mu G_1$, leading to
\begin{equation}\label{eq:EOM-NormalModes}
    (\omega-kv)\psi_v=\mu G_v(\psi_0-v\psi_1),
\end{equation}
where $k$ is always taken to be real, representing a spatial Fourier mode of the flavor field, whereas $\omega=\wR+i\wI$ may have a nonvanishing imaginary part, allowing for exponential growth or damping. 

We need to distinguish two cases. For modes with a 
phase velocity $\omega/k$ that is {\em superluminal}, i.e., $|\wR/k|>1$, the factor $(\omega-vk)$ never vanishes for any $-1\leq v\leq+1$, whether or not $\omega$ is complex, and we may write
\begin{equation}\label{eq:definition_eigenstates}
    \psi_v=\frac{\mu G_v}{\omega-kv}(\psi_0-v\psi_1).
\end{equation}
Inserting this form in Eq.~\eqref{eq:EOM-NormalModes} leads to the usual self-consistency condition
\begin{equation}\label{eq:dispersion_normal_modes}
    (\tilde{I}_0-1)(\tilde{I}_2+1)-\tilde{I}_1^2=0,
\end{equation}
where we introduce the integrals
\begin{equation}\label{eq:I-tilde-definition}
    \tilde{I}_n=\mu\int_{-1}^{+1}\!dv \frac{G_v v^n}{\omega-kv}.
\end{equation}
Notice that our definition of $\psi_v$ as the lower-left piece of the density matrix explains that the signs in $(\tilde I_0-1)$ and \smash{$(\tilde I_2+1)$} are opposite from some of the previous literature, e.g.~Eq.~(34) of Ref.~\cite{Yi:2019hrp}, but one finds the same $\omega(k)$ that fulfills Eq.~\eqref{eq:dispersion_normal_modes}, which is the dispersion relation for superluminal axisymmetric collective modes. 

We mention in passing that one can follow analogous steps for modes for which the initial condition is axisymmetric, yet the solutions break this symmetry. In other words, the spectrum $G_v$ still depends only on $v=\cos\theta$, whereas $\psi_\bv$ depends also on the azimuth angle $\phi$. For these modes, the dispersion relation has the simpler form
\begin{equation}\label{eq:dispersion-axi-breaking}
    \tilde{I}_0-\tilde{I}_2+2=0.
\end{equation}
The analytic properties of these modes are somewhat simpler than the axisymmetric ones, because their dispersion relation is linear in the integrals $\tilde{I}_n$ rather than quadratic. Here we do not further discuss them, since they are unrelated to the flavor pendulum.

For {\em subluminal\/} modes, i.e., $|\wR/k|<1$, the factor $(\wR-vk)$ vanishes for some $v$, corresponding to neutrinos with a velocity along the $z$-direction that is on resonance with the wave. If $\omega$ is complex, we can still proceed as before and obtain the dispersion relation for subluminal modes with a growing and damped mode $\omega=\wR\pm i|\wI|$. If the only goal is to find the exponentially growing modes of the system, there is no further difficulty.

However, to expand a given initial condition in a complete set of linearly independent eigenfunctions, one needs to note that in addition, there are real-valued modes with singular eigenfunctions that cannot be represented in the form of Eq.~\eqref{eq:definition_eigenstates}. For a given $k$, any frequency $-|k|<\omega<+|k|$ is possible for them---there is no dispersion law. In the context of flavor waves, these ``noncollective'' modes were first discussed in Ref.~\cite{Capozzi:2019lso}, whereas in plasma physics, they were discovered decades ago by van Kampen \cite{VanKampen:1955wh} and today are known as Case-van Kampen modes \cite{VanKampen:1955wh, Case:1959, Sagan:1993es}. In the noninteracting limit of $\mu\to0$, these are the only remaining modes and then correspond to individual neutrino waves that drift kinematically without being affected by other neutrinos.

\subsection{Superluminal real-valued modes}

Propagating modes $(\omega,k)$ with purely real $\omega$ and $k$ can only exist outside of the light cone, i.e., with superluminal phase velocity $|u|=|\omega/k|>1$. In this case, one can derive an explicit parametric expression for the dispersion relation \cite{Izaguirre:2016gsx}. We may express the integrals in Eq.~\eqref{eq:I-tilde-definition} in the form $\tilde{I}_n=(\mu/k)\,J_n$ with
\begin{equation}
\label{eq:jn}
    J_n=\int_{-1}^{+1}\!dv\,\frac{G_v v^n}{u-v},
\end{equation}
which is a function of the phase velocity $u$. These expressions are well defined because for $|u|>1$, the integrand has no singularity. The dispersion relation Eq.~\eqref{eq:dispersion_normal_modes} then reads
\begin{equation}
    \left(J_0-k\right) \left(J_2+k\right)=J_1^2
\end{equation}
which is a quadratic equation for $k$ with the solutions
\begin{equation}\label{eq:k_function_u}
    k=\frac{J_0-J_2}{2}
    \pm\sqrt{\left(\frac{J_0+J_2}{2}\right)^2-J_1^2}.
\end{equation}
Here and henceforth we assume that $\omega$ and $k$ are given in units of $\mu$, so we have set $k/\mu\to k$. For that range of $1<|u|<\infty$, when the expression under the root is positive, one finds the \edit{real-valued} branches $k_{1,2}(u)$ and $\omega_{1,2}(u)=u\,k_{1,2}(u)$, representing \edit{real-valued} branches $(\omega,k)$ of the dispersion relation in parametric form. These equations are more generally valid for any complex value of $u$, including the subluminal ones, but in such cases they are not guaranteed to lead to a real wavenumber, since the integrals $J_n$ (completed with the $i\epsilon$ prescription, see below), are generally complex. For superluminal modes, the integrals $J_n$ are real, so the wavenumber will be real provided that the argument of the square root is positive.

For the axially breaking modes, an analogous, simpler expression can be found~\cite{Yi:2019hrp, Fiorillo:2024uki}
\begin{equation}
    k=\frac{J_2-J_0}{2}.
\end{equation}
Given that there is no square root, physical modes with real wavenumbers exist for any value of the superluminal phase velocity $u$.

\subsection{Discrete modes}

Numerical simulations use a discrete representation of phase space, here of the velocities $v_i$ ($i=1,\dots,N$). For a given $k$, one solves a set of $N$ equations, i.e., the eigenvalues $\omega_i(k)$ of an $N{\times}N$ matrix. Most of them will be real, but there can be complex ones which are assured if the spectrum $G_{v_i}$ has a single crossing~\cite{Capozzi:2019lso, Fiorillo:2024uki}. As a function of $k$, these emerge as the coalescence of two real-valued modes because the total number $N$ of solutions remains the same. So a complex branch as a function of $k$ always ends in a critical point of two coalescing real-valued modes, which under the light cone are two Case-van Kampen modes in the continuum limit \cite{Capozzi:2019lso}, whereas above the light cone would be two real-valued collective modes \cite{Yi:2019hrp}. For a given $G_v$, one can solve the discrete counterpart and get a first glimpse of the stable and unstable collective modes in terms of the functions $\omega_i(k)$.

\subsection{Subluminal physical modes}

Above the light cone, the stable and unstable normal modes found by Eq.~\eqref{eq:dispersion_normal_modes} provide all the information that we seek on the level of the dispersion relation. Under the light cone, the situation is more complicated. The subluminal unstable growing modes are also physical and if this is all one wants to know, again one has all the information. On the other hand, these are not all the physical modes. In the discrete case, one finds $N$ branches of the dispersion relation, most of them real, but not physical, because the medium does not fall into discrete modes. In the same way, in the continuum case, the singular Case-van Kampen modes are also not physical because one cannot excite a singular wave function. Physical waves are superpositions of normal modes.

In plasma physics, this question was resolved a long time ago first by Landau~\cite{Landau:1946jc} and recently introduced to the flavor community by two of us \cite{Fiorillo:2023mze, Fiorillo:2024bzm, Fiorillo:2024uki}. In a nonrelativistic isotropic plasma, the physically oscillating modes are Langmuir waves, also known as longitudinal plasmons, which physically represent a collective oscillation of the electrons against the positively charged background of heavy ions. However, some electrons always have the same speed as the phase velocity of the plasmon and are in this sense on resonance with the wave. They drain energy by Cherenkov absorption, an effect widely known as Landau damping. Conversely, if the electron velocity distribution is not isotropic but exhibits, for example, the celebrated bump-on-tail shape, the plasma wave grows by Cherenkov absorption from the resonant electrons and thus to an instability by inverse Landau damping
\cite{thorne2017modern}. The exact condition on the electron velocity distribution for this to happen is the Nyquist criterion derived by Oliver Penrose \cite{Penrose:1960}.

The lesson is that physical plasmons are either Landau damped or grow. How does this picture translate to flavor waves? As we have reviewed in Ref.~\cite{Fiorillo:2024bzm}, this leads to a surprisingly simple conclusion: the asymptotic collective behavior at late times is determined by eigenfrequencies which solve a dispersion relation of the same form as
Eq.~\eqref{eq:dispersion_normal_modes}
\begin{equation}\label{eq:landau_dispersion}
    (I_0-1)(I_2+1)-I_1^2=0.
\end{equation}
The only difference is that the integrals are modified to
\begin{equation}\label{eq:landau_dispersion_integrals}
    I_n=\int_{-1}^{+1}\!dv\,\frac{G_v v^n}{\omega-kv+i\epsilon}\,,
\end{equation}
including an infinitesimal shift of the position of the pole in the denominator. For real frequencies, the $i\epsilon$ prescription can be understood from simple physical arguments; a small imaginary part corresponds to the perturbation slowly inserted from $t\to -\infty$, so this prescription ensures that we only obtain the modes that are actually visible in the far future and respect causality. Equivalently, we can imagine $\epsilon$ to be a small collisional damping inserted by hand to regularize the dynamics. 

 The $+i\epsilon$ prescription requires that the integral over $v$ must be done along a path in the complex plane passing \textit{below} the pole $v=\omega/k$. The modification is analogous for the axi-breaking modes of Eq.~\eqref{eq:dispersion-axi-breaking}. Hence, the  solutions of Eqs.~\eqref{eq:dispersion_normal_modes} and~\eqref{eq:landau_dispersion} coincide when $\mathrm{Im}\,\omega>0$, since in that case the integration along the real axis always passes below the pole. They also coincide for superluminal modes, since in that case the integrand simply vanishes below the pole. However, crucially, for damped subluminal modes ($\mathrm{Im}\,\omega<0$) the two dispersion relations differ. In the normal-mode approach, damped modes always and only appear as complex conjugates of unstable modes. In the Landau approach, this applies only to superluminal modes. For subluminal modes, on the other hand, an unstable mode is not accompanied by a complex conjugate damped mode. Rather, novel damped modes appear which are not normal modes, but Landau-damped oscillating modes. The reason they are not normal modes is because it is not the individual $\psi_v$ that are being damped; rather, the individual $\psi_v$ remain of the same amplitude, but the collective behavior, as measured, e.g., by $\psi_0=\int dv \psi_v$, is exponentially damped due to phase mixing. Yet, since we are usually interested only in collective behavior, and not in the flavor of individual velocity modes, Landau damping is a physical effect.

To summarize, subluminal normal modes with nonvanishing growth rate are accompanied by a complex conjugate damped wave, and the pair emerges from the coalescence of two singular Case-van Kampen modes. In the Landau picture, under the light cone a collective mode with given $k$ is either a Landau-damped oscillating one, or an exponentially growing (unstable) one without damped counterpart. For a given $k$, growth or damping are alternatives, not simultaneous forms of evolution. In this picture, an unstable branch under the light cone does not simply end as a function of $k$, but rather continues as a Landau-damped mode.

The discussion here has been somewhat formal and gives the gist of what we have derived in more detail in Refs.~\cite{Fiorillo:2023mze, Fiorillo:2024bzm, Fiorillo:2024uki}. In Sec.~\ref{sec:Example}, we will underpin this discussion with explicit examples.

\subsection{Homogeneous mode}\label{sec:homogeneous_mode}

An often-studied case is that of a homogeneous mode ($K=0$), seemingly the simplest possible setup, namely a perfectly homogeneous neutrino gas with an axisymmetric perturbation. The dynamics is periodic, due to a large set of conservation laws, the so-called Gaudin invariants~\cite{Pehlivan:2011hp, Fiorillo:2023mze, Fiorillo:2023hlk}, and actually it can be mapped to that of a pendulum~\cite{Johns:2019izj, Padilla-Gay:2021haz}, the fast flavor pendulum. An explicit form of the mapping is given in Ref.~\cite{Fiorillo:2023hlk}. The special value of wavenumber $K=0$, corresponding to $k=G_1$, leads to a very simple form for the dispersion relation for longitudinal modes:
{\color{black}
\begin{equation}
\label{eq:goldstone}
    \left(\omega-G_0\right)\int_{-1}^{+1}\!dv\,\frac{G_v v}{\omega-G_1 v+i\epsilon}=0\,.
\end{equation}
This dispersion relation always admits a trivial solution $\omega=G_0$ and therefore $\Omega=0$, which describes a homogeneous neutrino gas that just advects with uniform velocity. Such a solution is possible because neutrino self-interactions locally preserve lepton number. In this sense, $(K=0,\Omega=0)$  may be called a Goldstone mode: a zero-energy excitation whose existence is guaranteed by conservation laws.

Factoring out this term, the remaining solutions satisfy $I_1=0$, or explicitly:
\begin{equation}\label{eq:dispersion_homogeneous_mode}
\int_{-1}^{+1}\!dv\,\frac{G_v v}{\omega-G_1 v+i\epsilon}=0\,.
\end{equation}
This is the dispersion law for the fast flavor pendulum. Can we determine from the properties of $G_v$ if there are unstable solutions without having to solve this equation? This is our main question.
}

We stress that this simplified dispersion relation, as well as any conclusion about the regularity of the pendular behavior even in the nonlinear regime, applies \textit{only} to this perfectly homogeneous and axisymmetric case. The conservation laws associated with the pendular behavior are strictly protected by these symmetries, and any small breaking of them destroys its regularity. Moreover, a matter background with a net flow {\color{black}breaks the conservation of the Gaudin invariants and destroys the regular behavior~\cite{Fiorillo:2023hlk},} which therefore is probably hopeless to witness in a realistic environment.

Still, given the attention that the fast flavor pendulum has commanded in the past, including by some of the present authors, as an explicit prototypical case of a nonlinear collective system, it is worth considering its stability condition, especially to connect with the more modern viewpoint of the instability as a kinetic phenomenon.

\section{Nyquist criterion}
\label{sec:Nyquist}
The general problem of determining whether an equation admits unstable solutions, with a positive imaginary part, can usually be tackled by a simple theorem of complex analysis, that we will briefly review here. This approach has led in plasma physics to the so-called Nyquist criterion, to determine whether a plasma with a given energy and velocity distribution for the electrons possesses an instability, without necessarily solving its dispersion relation \cite{schKT, Penrose:1960}. Motivated by this analogy, two of us have proposed a similar Nyquist criterion to determine the existence of the fast pendulum instability, \edit{i.e., of an unstable homogeneous mode\footnote{\edit{If $v_c=0$, then the homogeneous mode is always stable. The Nyquist criterion states that unstable modes necessarily lie in an interval $(k_-,k_+)$, such that $u(k_\pm)=v_c$. To find the wavenumbers $k_\pm$, one can insert $u=v_c=0$ in Eq.~\eqref{eq:jn}, which shows that $J_1=-G_0$ and $J_2=-G_1$. Equation~\eqref{eq:k_function_u} then implies that the physical wavenumbers $K_\pm =k_\pm -G_1$ both have the same sign, so $K=0$ is excluded from the unstable interval.}}~\cite{Fiorillo:2023hlk}:
\begin{equation}\label{eq:OldNyquist}
    \frac{G_1}{G_0 \vc}<0
    \quad\hbox{and}\quad
    \int_{-1}^{+1}dv\,\frac{v G_v}{G_0(v-\vc)}<0\,,
\end{equation}
where the crossing is at $\vc$, i.e., $G(\vc)=0$. Contrary to this previous finding, this condition is only necessary: as such, it is more restrictive than the condition of a single crossing, but it is not always sufficient}. It turns out that the fast flavor instability is more complicated than the standard case of nonrelativistic plasmas. Therefore, after reviewing the original formulation, we will explain how the Nyquist criterion fails in the flavor context, largely deriving from the finite range $-1\leq v\leq+1$ of velocity integration. Somewhat counter-intuitively, in a nonrelativistic plasma, the integration is over the range $-\infty<v<+\infty$, there is no limiting velocity, although the high-velocity tails are exponentially suppressed for $v$ beyond a typical nonrelativistic electron velocity. Still, in nonrelativistic plasma, there are no superluminal modes.

\begin{figure}[ht]
    \centering
    \includegraphics[width=1.0\columnwidth]{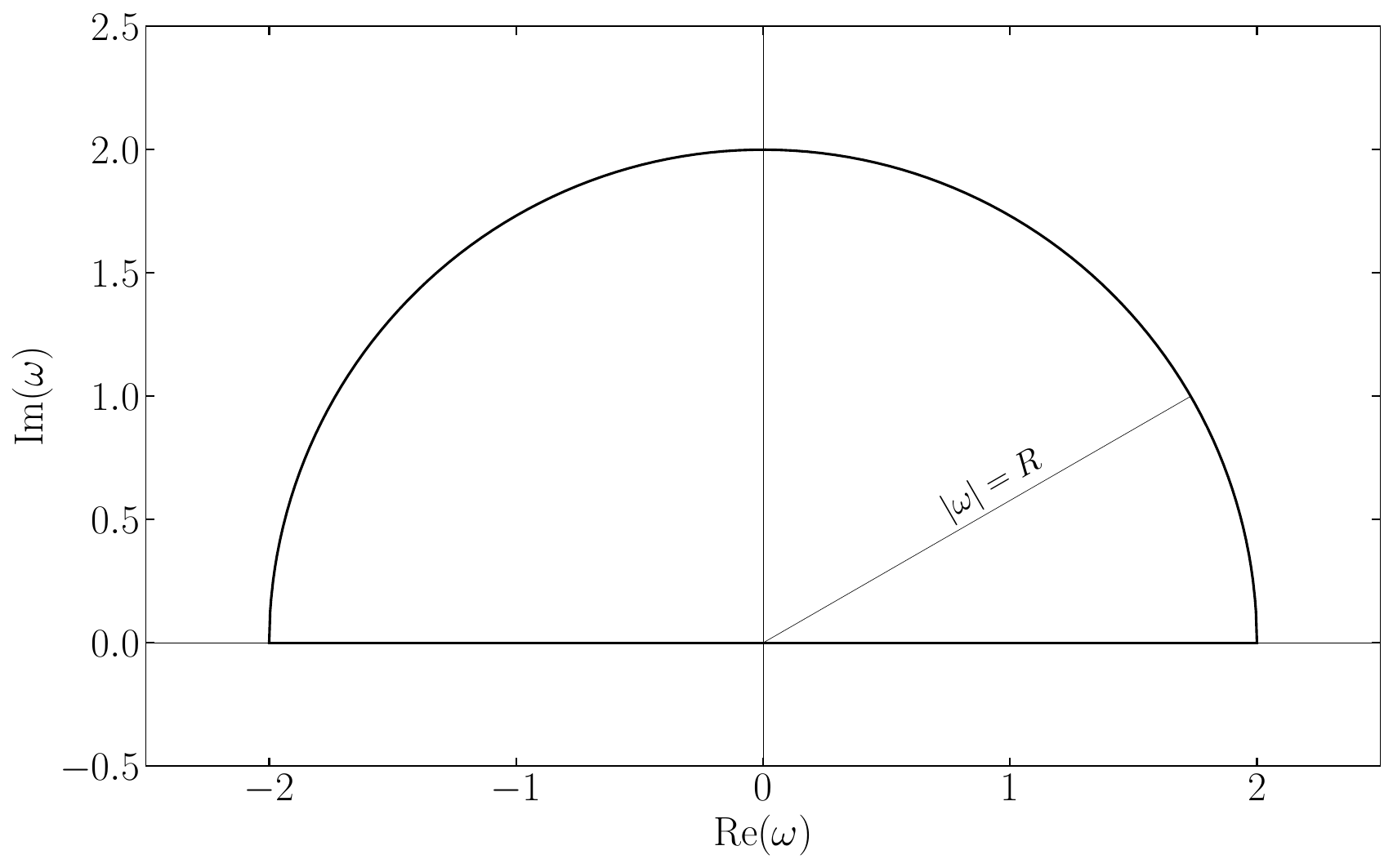}
    \caption{Integration path in the complex plane of frequency $\omega$ for the application of Nyquist's criterion.}
    \label{fig:Nyquist_plasma}
\end{figure}

Let us assume the dispersion relation is given by an equation of the form $\Phi(\omega)=0$ for a fixed value of the wavenumber $k$; for a homogeneous setup
($K=0$), we would choose $k=G_1$. We will initially assume $\Phi(\omega)$ to be an analytic function of $\omega$. The existence of solutions with $\mathrm{Im}(\omega)>0$, i.e., in the upper half-plane of the complex variable $\omega$, can now be determined by noting that
\begin{equation}
    \int_{-\infty}^{+\infty} \frac{\Phi'(\omega)}{\Phi(\omega)}\frac{d\omega}{2\pi i}+\lim_{R\to +\infty}\int_{0}^\pi \frac{\Phi'(R\,e^{i\varphi})}{\Phi(R\, e^{i\varphi})}\frac{R\,e^{i\varphi}d\varphi}{2\pi} =N,
\end{equation}
where $N$ is the number of zeroes of $\Phi(\omega)$ in the upper half-plane. This simple result follows directly from the theorem of residues, when the integrand function is integrated along a semicircle of radius $|\omega|=R$ surrounding the complex upper half-plane (Fig.~\ref{fig:Nyquist_plasma}).

\begin{figure*}
    \centering
    \includegraphics[width=0.5\textwidth]{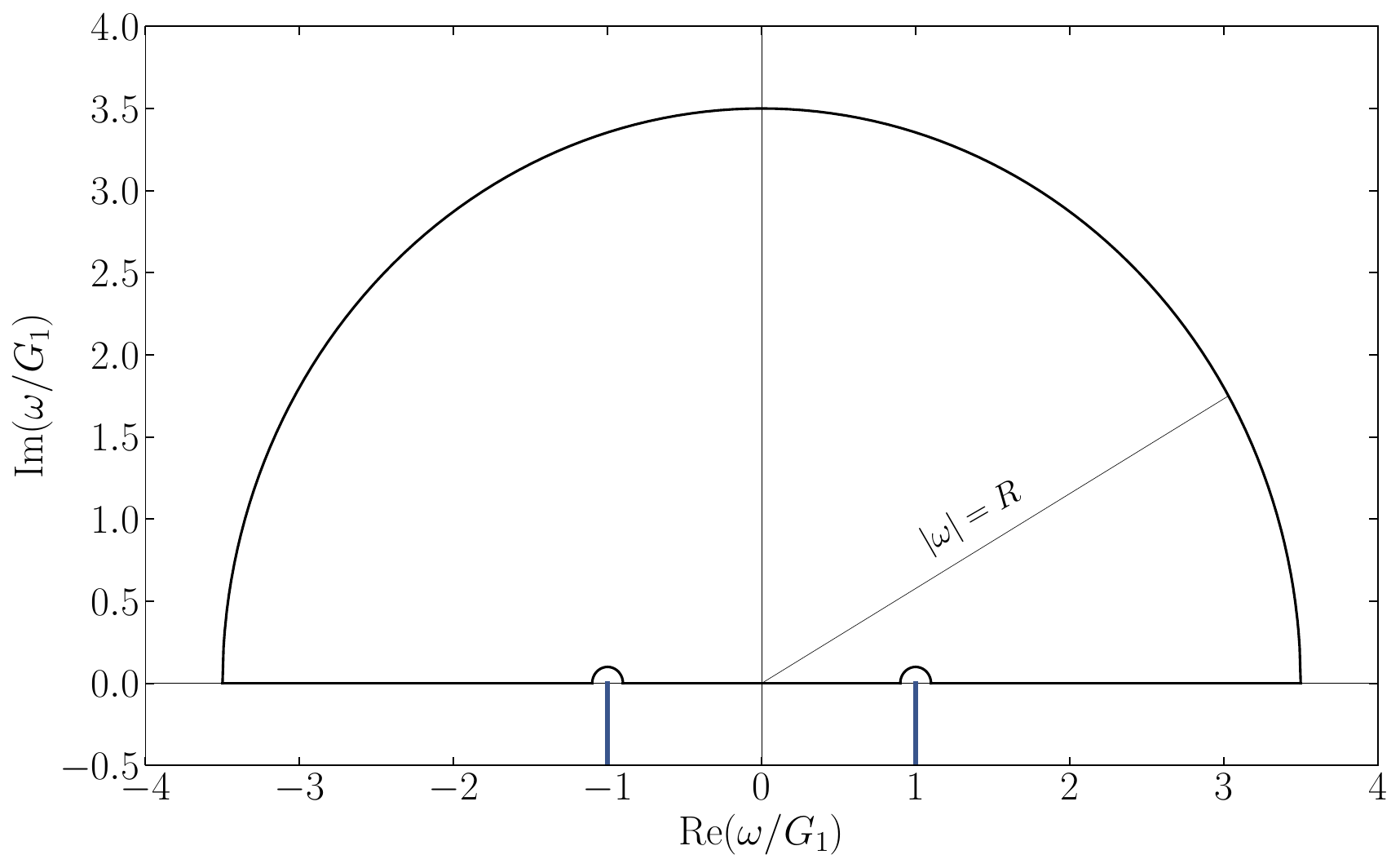}\includegraphics[width=0.5\textwidth]{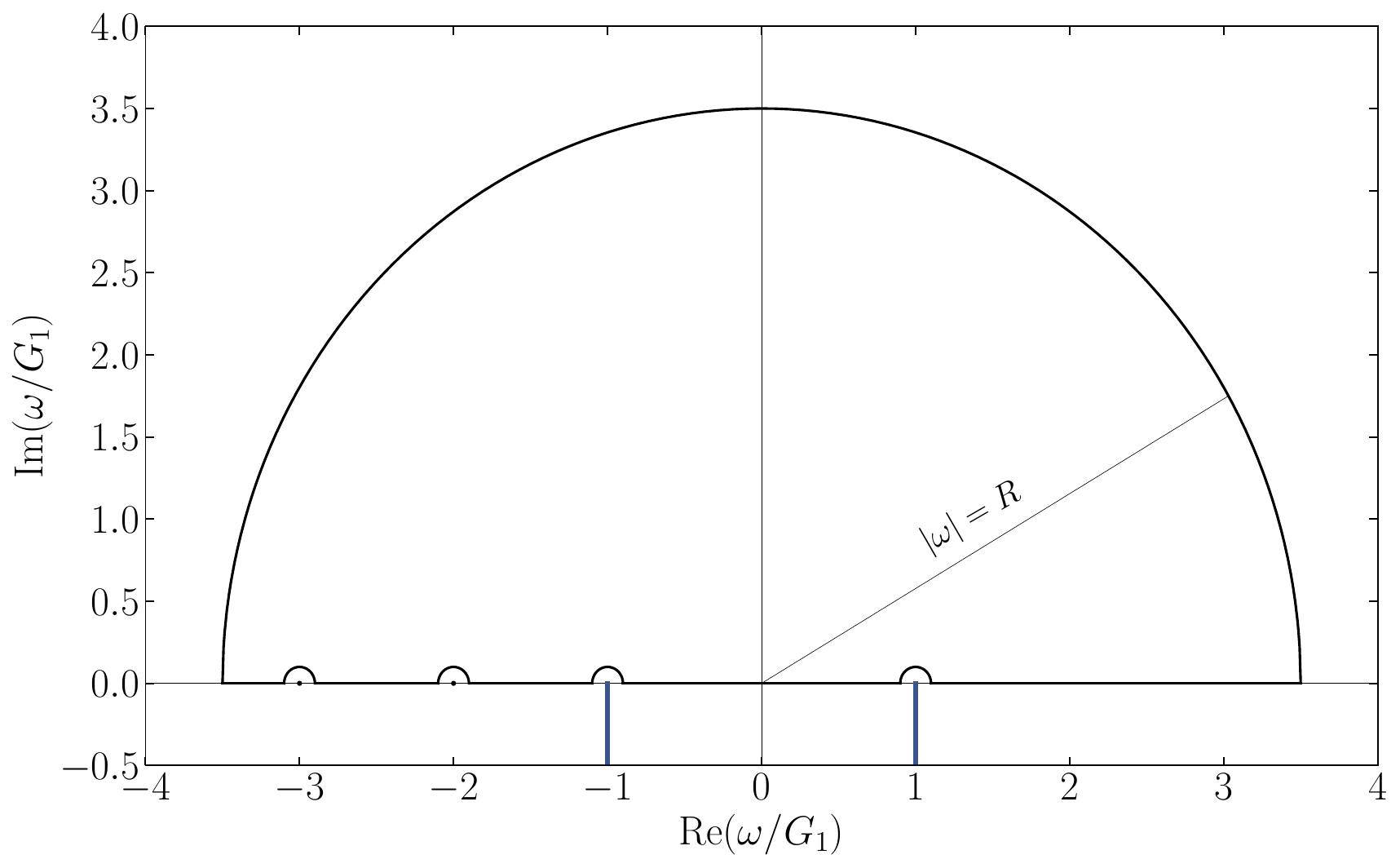}
    \caption{Integration path in the complex plane of the phase velocity $u$ for the application of Nyquist's criterion. The blue lines are branch cuts of the function $\Phi(u)$. We show the integration path for the case of no real-valued superluminal modes (left), and \edit{for a particular example with} two real-valued superluminal modes (right).}
    \label{fig:Nyquist_flavor}
\end{figure*}

In Ref.~\cite{Fiorillo:2023hlk}, we have applied this theorem to the special case of the homogeneous dispersion relation. Actually it is more convenient to do so by considering the dispersion relation not as a function of $\omega$, but rather of the phase velocity $u=\omega/k$, which for the homogeneous mode is $u=\omega/G_1$. This is a trivial rescaling, but since $G_1$ may be negative, it turns out to be more convenient in practice. The dispersion relation then reads
\begin{equation}
\label{eq:dr}
    \Phi(u)=\int_{-1}^{+1}\frac{G_v v}{u-v+i\epsilon}dv=0.
\end{equation}

The first difference compared to the plasma case, that was not sufficiently highlighted in the original derivation~\cite{Fiorillo:2023hlk}, is that the function $\Phi(u)$ is \textit{not} analytic. Since the integral is over the finite interval $-1<v<+1$, the imaginary part of $\Phi(u)$ clearly is discontinuous when $u$ passes through $\pm1$. The nature of this discontinuity, and the general analytic structure of the function, was discussed in Ref.~\cite{Fiorillo:2024bzm}. The function $\Phi(u)$ has two branch cuts on the lines $\mathrm{Re}(u)=\pm 1$ and $0\geq\mathrm{Im}(u)>-\infty$. Therefore, the integration path must be distorted as in Fig.~\ref{fig:Nyquist_flavor}, left panel. We will therefore imply by the notation $\int_{-\infty}^{+\infty} du$ the integral along the real axis thus deformed.

So, after replacing the asymptotic form of $\Phi(u)$ for $|u|\to+\infty$ to evaluate the integral along the large semicircle, we find the identity
\begin{equation}\label{eq:phase:accumulated}
    \int_{-\infty}^{+\infty} \frac{\Phi'(u)}{\Phi(u)}\frac{du}{2\pi i}=N+\frac{1}{2}.
\end{equation}
The final step to apply Nyquist's criterion is to notice that the first integral can be rewritten as
\begin{equation}
    \int_{-\infty}^{+\infty}\frac{\Phi'(u)}{\Phi(u)}\frac{du}{2\pi i}=\frac{1}{2\pi i}\log\left[\frac{\Phi(u\to+\infty)}{\Phi(u\to-\infty)}\right],
\end{equation}
which is simply the phase accumulated by $\Phi(u)$ as $u$ runs along the real axis. This phase can be determined by a simple graphical approach; the (generally complex) function $\Phi(u)$ for real $u$ draws a trajectory in the complex plane, and the number of times this trajectory wraps around the origin -- the winding number $W$ -- gives the phase accumulated in units of $2\pi$. The $1/2$ term on the right-hand side simply accounts for the fact that the function $\Phi(u)\to G_1/u$ at $\pm \infty$ changes sign, and therefore does not enter the definition of the winding number. Notice that the integral along the two small semicircles around $u=\pm 1$, while conceptually necessary, in practice vanishes, because $\phi\to\infty$ close to the beginning of the branching line, while $\phi'\to 0$.
So, if the trajectory does not wrap around the origin, there will be no unstable solution, and more generally $W=N$. Based on this geometrical argument, in Ref.~\cite{Fiorillo:2023hlk} the criterion Eq.~\eqref{eq:OldNyquist} was found for the existence of an unstable homogeneous solution for a single-crossed angular distribution.

However, there is one simple way in which the above theorem fails. We have assumed that $\Phi(u)$ never vanishes on the integration path. For subluminal $|u|<1$ this is essentially always true, since $\mathrm{Im}[\Phi(u)]$ can only vanish if $u=0$ or $u=\vc$ (the crossing velocity of $G_v$). The vanishing of $\mathrm{Re}[\Phi(u)]$ at exactly one of these two points is just the condition separating the regimes in which an unstable mode does or does not exist. However, it was overlooked that $\Phi(u)$ might vanish for real-valued superluminal $u$. In this case, $\mathrm{Im}[\Phi(u)]$ vanishes identically, so that $\mathrm{Re}[\Phi(u)]$ might in principle vanish at any superluminal value of $u$. If some such points exist, corresponding to real-valued superluminal modes that solve the dispersion relation, the integration path must be further modified as in the right panel of Fig.~\ref{fig:Nyquist_flavor} to pass over the poles. At each small semicircle, $\Phi(u)$ changes sign and therefore accumulates a phase $e^{-i\pi}$, so that ultimately if there are $N_s$ superluminal modes, the winding number $W$ will no longer equal the number of unstable modes $N$, but rather $N+N_s/2$, since the left-hand side of Eq.~\eqref{eq:phase:accumulated} accumulates a contribution $-1/2$ for each zero on the real axis. 

Therefore, the Nyquist criterion still applies, but only by accounting for this modification. Essentially, $\Phi(u)$ accumulates a phase both when $u$ is subluminal and when it is superluminal; our previous version only accounted for the former. Equation~\eqref{eq:OldNyquist} still remains true as a necessary criterion, since in order to have $N>0$, we must have $W>N_s/2$ and therefore $W>0$. {\color{black}For a single-crossed distribution, $N_s$ is always even because $\Phi(u)\to G_1/u$ for $u\to \infty$, so it has opposite signs at $u\to\pm \infty$. Instead, for $u\to +1^+$ we have $\Phi(u)\to G(1)\log\left[2/(u-1)\right]$, while for $u\to -1^-$ we have $\Phi(u)\to G(-1)\log\left[2/(-u-1)\right]$, so it also changes sign at these two points. It follows that the number of times that $\Phi(u)$ can change sign in the superluminal intervals is necessarily even, i.e., $\Phi(u)$ has an even number $N_s$ of superluminal zeroes. Hence, the number of superluminal modes is odd, and is equal to $N_s+1$, due to the additional Goldstone mode whose existence we have noted in Sec.~\ref{sec:homogeneous_mode}.}

So we finally reach a generalized strategy to apply the Nyquist criterion to a given angular distribution. We first check what is its winding number $W$ in the subluminal range $-1<u<1$. For a single-crossed angular distribution, the criterion from Ref.~\cite{Fiorillo:2023hlk} allows us to determine directly whether $W=0$ or $W=1$. If $W=0$, and therefore Eq.~\eqref{eq:OldNyquist} is violated, there is no instability. If $W=1$, we can graphically plot $\Phi(u)$ for $u<-1$ and $u>1$, and count the number of zeroes. If there are none, then there is an instability, while if there are two zeroes, there is no instability. More generally, the number of unstable modes for any given angular distribution, even with more than one crossing, is given by $N=W-N_s/2$.

While these arguments pertain only to the homogeneous mode, the same conclusions can be reached on more general grounds considering how the instability arises as a function of wavenumber. In Sec.~\ref{sec:General}, we will provide this complementary discussion, showing that indeed it leads to the same results. \edit{However, we first examine the branches of the dispersion relation of a few single-crossed angular distributions in Sec.~\ref{sec:Example}, and apply the corrected Nyquist criterion to count the number of unstable modes.}

\section{Illustrating example}
\label{sec:Example}

\begin{figure}[h]
    \centering
    \includegraphics[width=0.98\columnwidth]{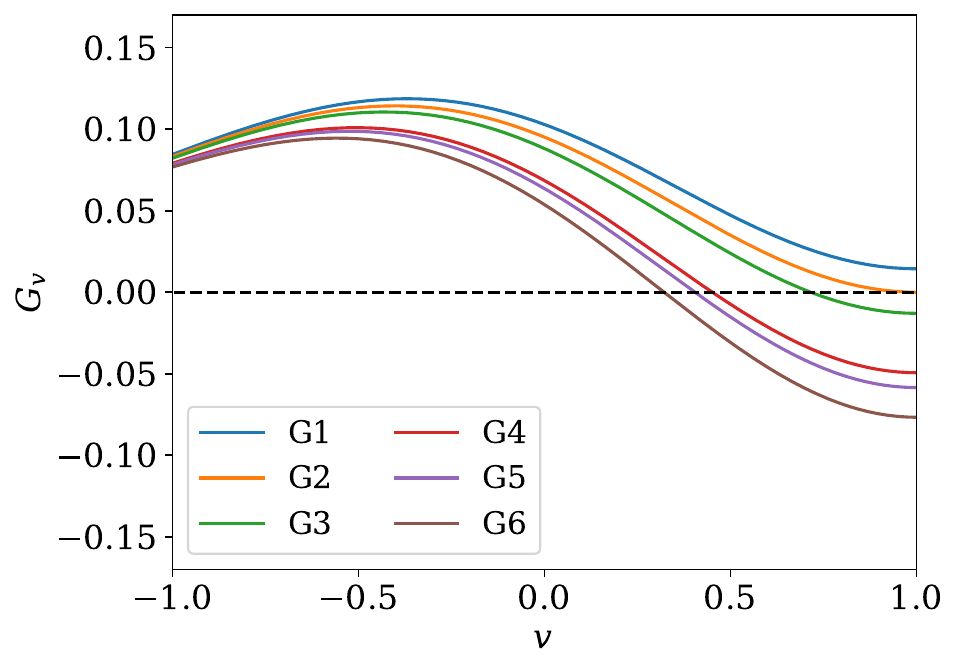}
    \vskip-8pt
    \caption{Angular spectra $G_v$ used for our generic examples, parametrized according to Eq.~\eqref{eq:ang_spectrum} and the $a$ values in Table~\ref{tab:ex_params}.}
    \label{fig:spectra}
\end{figure}

To illustrate the generic cases of dispersion of an axially symmetric system, we are inspired by the angular spectra $G_v$ presented in Ref.~\cite{Yi:2019hrp}. The $\nu_e$ and $\overline\nu_e$ distributions are each taken to be Gaussians of the form $a e^{-(v-1)^2/2b^2}$, such that overall
\begin{equation}
\label{eq:ang_spectrum}
    G_v=\frac{e^{-(v-1)^2/2b_{\nu_e}^2}}{N_{\nu_e}}-a\,\frac{e^{-(v-1)^2/2b_{\overline\nu_e}^2}}{N_{\overline\nu_e}},
\end{equation}
where $N_l=\int_{-1}^{+1} dv e^{-(v-1)^2/2b_l^2}$. We use the parameters
$b_{\nu_e}=1.1$ and $b_{\overline\nu_e}=0.9$, and we vary $a$ to change the overall shape of $G(v)$; see the specific values in Table~\ref{tab:ex_params}. Small $a$ values produce no angular crossing, i.e., $G_v>0$ everywhere; larger ones result in a single crossing at some $v=v_c$. Initially, the ``flipped'' part of the distribution is located at $v>0$ and grows larger with $a$, as shown in Fig.~\ref{fig:spectra}. However, as $a$ increases further, the spectrum becomes negative in most of the angular interval: at that point, the $v<0$ side of the distribution is the one with a small flipped region, which would eventually disappear for large enough values of $a$. Thus, Eq.~\eqref{eq:ang_spectrum} sweeps over the generic cases of a single-crossed spectrum. 

\begin{table}[h]
\caption{Properties of the angular distributions in Fig.~\ref{fig:spectra}, where numerically $a_2=N_{\overline\nu_e}/N_{\nu_e}=0.855769$. }\label{tab:ex_params}
\vskip2pt
\edit{\begin{tabular*}{\columnwidth}{@{\extracolsep{\fill}}lllllll}
    \hline\hline
    Case&G1&G2&G3&G4&G5&G6\\
    $a$& 0.84 & $\displaystyle\frac{N_{\overline\nu_e}}{N_{\nu_e}}$ &  0.87 & 0.91 &  0.92 & 0.94\\
    $G_0$ & $0.16$ & $0.14$ & $0.13 $&$ 0.090$ & $0.080 $& $0.060$  \\
    $-G_1$ & $0.035$ & $0.040$ & $0.045$ & $0.058$ & $0.061$ & $0.068$\\
    $W$   & 0 & 0 & 1 & 1 & 1 & 1 \\
    $N_s$ & 0 & 0 & 0 & 0 & 0 & 0 \\
    $N$   & 0 & 0 & 1 & 1 & 1 & 1 \\

    \hline
\end{tabular*}}
\end{table}

\exclude{\color{teal} Regarding how to parameterize a family of curves in terms of parameter which might be roughly connected with the instability properties, one possibility is a slight reparameterization of the NBI Gaussian beam choice
\begin{equation}
    G_v=\frac{\exp\left[-\frac{(1-v_c)^2}{2d^2}\right]-\exp\left[-\frac{(1-v)^2}{2d^2}\right]}{1-\exp\left[-\frac{2}{d^2}\right]},
\end{equation}
where $d$ measures the width of the Gaussian beam, and $v_c$ is the crossing. This is a minimal parameterization -- a third parameter would simply be a scaling factor to amplify or reduce the amplitude of $G_v$. Right now, this is normalized such that $G_1-G_{-1}=1$. For making plots, it could also be useful to show things in terms of $v_c$, the crossing velocity, which means how wide or narrow the "flipped" region is (i.e. the region where $G_v$ changes sign), and $G_1$, which measures how deep the flipped region is. So these two quantities give a measure of how strong the instability is expected to be.}

\begin{figure*}
\centering
\includegraphics[width=\textwidth]{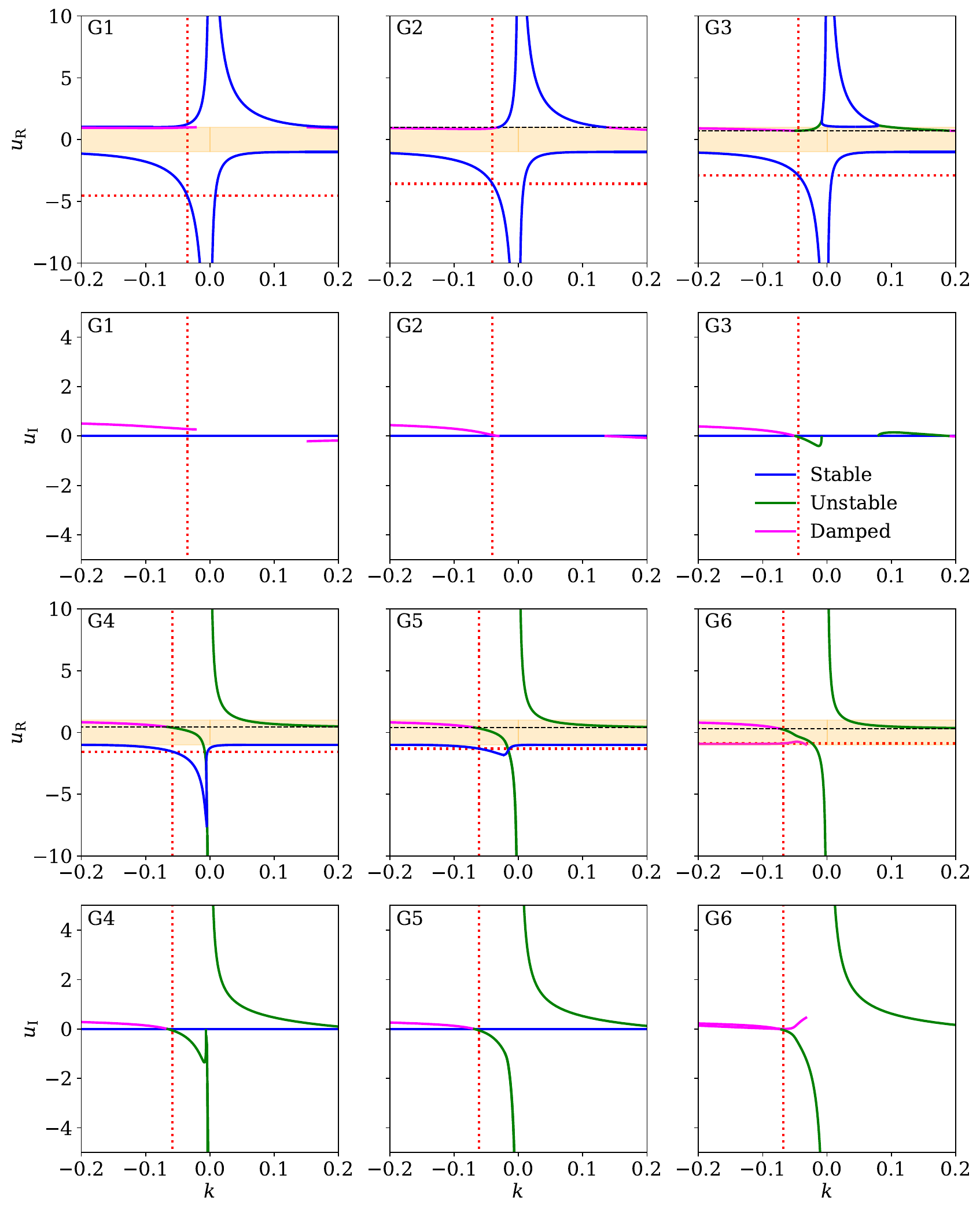}
\caption{Branches of the dispersion relation for our reference angular spectra. {\color{black}The vertical dotted red line marks $k=G_1$, i.e., the homogeneous mode for each angular distribution.} In the first and third row of panels, the vertical axis displays $\uR$, the real part of the phase velocity. {\color{black} The horizontal dotted red line shows $\uR =G_0/G_1$, {\color{black}i.e., the intersection of the dotted red lines marks the Goldstone mode.} 
The orange band covers subluminal modes ($|\uR|<1$) and is delimited by the light cones $\uR=\pm1$.} The dashed black line marks the crossing velocity, $\uR=\vc$. The blue  lines correspond to $\omega\in \mathbb{R}$ and have been calculated using Eq.~\eqref{eq:k_function_u}. In the second and forth rows, the imaginary part $\uI$ is shown.  The green (magenta) lines are branches with $\wI >0$ ($\wI <0$). {\color{black}Even in G6, the Goldstone mode is stable ($\wI=0$), being a single stable point on the Landau-damped branch.}}
\label{fig:DR_branches}
\end{figure*}

\begin{figure*}
\centering
\includegraphics[width=\textwidth]{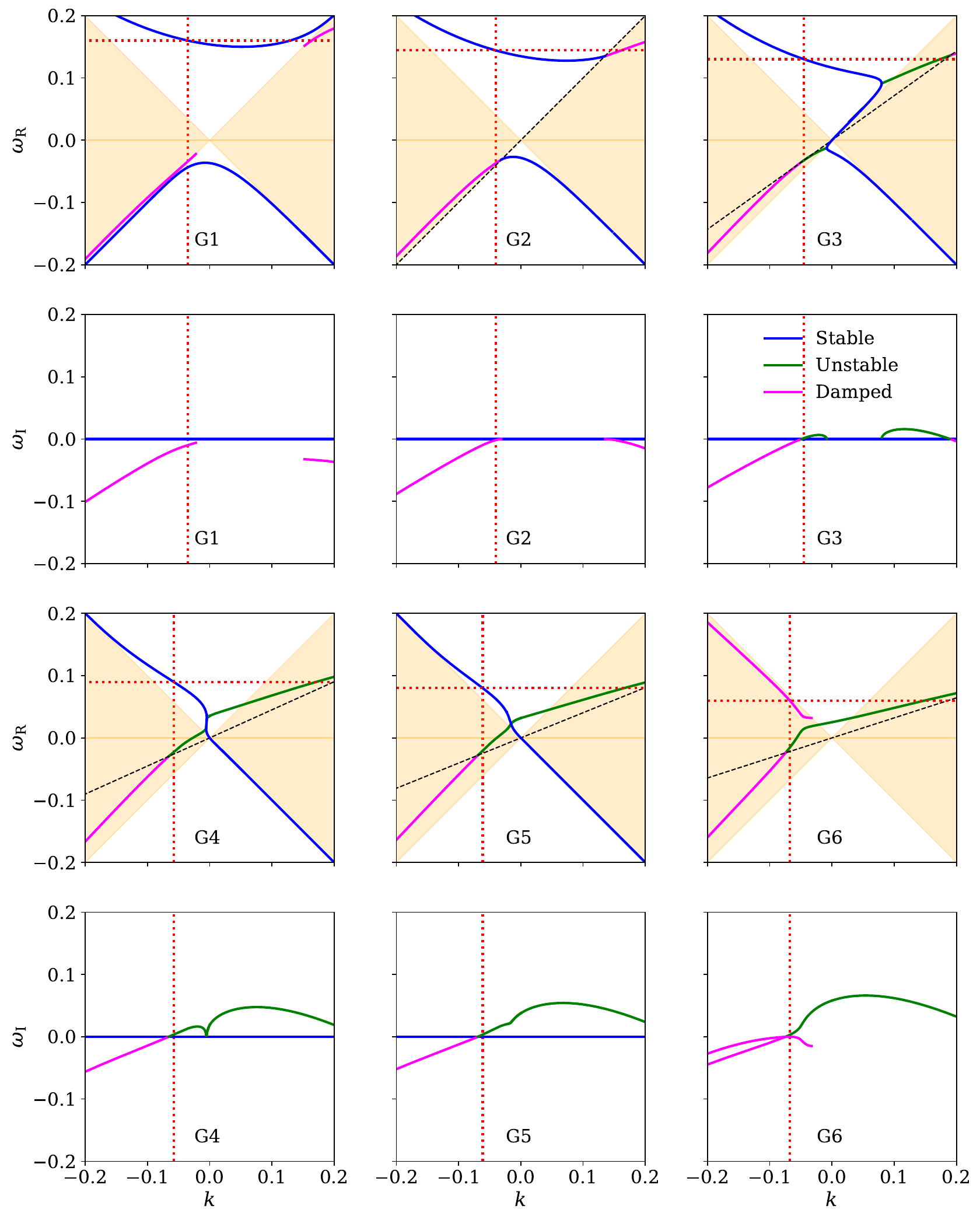}
\caption{Same as Fig.~\ref{fig:DR_branches}, but now the vertical axis
shows the real (imaginary) part of the angular frequency, $\wR$ ($\wI$).}
\label{fig:DR_branches_w}
\end{figure*}

The branches of the dispersion relation with $k\in \mathbb{R}$ and $\omega \in \mathbb{R}$ or $\mathbb{C}$ are shown in Figs.~\ref{fig:DR_branches}
and \ref{fig:DR_branches_w} for these distributions. In agreement with earlier studies~\cite{Yi:2019hrp}, the results can be generically divided into the following categories: 

\begin{itemize}
    \item[{\bf 1.}] {\bf No crossing (G1, G2)}.\\ 
    If $G_v$ has no crossing, there is no instability. There are two \edit{real-valued} branches of superluminal modes $u(k)$, asymptoting as \hbox{$u(k\to\pm\infty)\to\pm 1$} to the light cone. In addition, there are subluminal branches of Landau-damped modes, not identified in Ref.~\cite{Yi:2019hrp}, who only considered normal modes. It is only for $u=+1$ that both a real and a Landau-damped branch appear close to the light cone, but not for $u=-1$, an asymmetry that is explained by $G_v$ having ``almost a crossing'' at $u=+1$. The Landau-damped branches develop in anticipation, so to speak, of the appearance of a crossing. We will see shortly that these branches evolve into the more familiar unstable modes once a crossing has developed.
    
    For case G2, when $G_v=0$ exactly at $v=+1$, the Landau-damped branches characteristically do not stop at $\mathrm{Re}(u)=1$, but instead smoothly merge with the real-valued superluminal branches. There is no discontinuity in $G_v$ because it vanishes at $v=+1$ (although there is one in the derivative), and therefore the branch cut mentioned above vanishes, allowing for a continuous transition from the real to the Landau-damped modes.
    
    \item[{\bf 2.}] {\bf Shallow crossing (G3)}. \\ 
    When a small part of the spectrum changes sign, two complex branches appear: each of them begins under the light cone as a Landau-damped mode, crosses $u=\vc$, and terminates on a critical point of the \edit{real-valued} branch. The latter now passes through the origin: \edit{this creates a triangular feature in the upper half of the $u(k)$ plot (Fig. \ref{fig:DR_branches}, top right), which appears as a characteristic S-shape in the $\omega (k)$ diagram (Fig. \ref{fig:DR_branches_w}, top right).  Consequently, a single value of $k$ may have three associated frequencies}. 
    
    The real-valued modes with negative $\uR$ are essentially not affected by the instability. This is particularly clear from the viewpoint of the instability as a resonant interaction between flavor waves and neutrinos~\cite{Fiorillo:2024bzm}: since the unstable flavor waves are resonant with neutrinos in the ``flipped'' region, which in this case is very close to $v=+1$, nothing new happens in the region close to $u=-1$. 
    
    We also notice that, in principle, the real-valued modes with $\uR>1$ can exhibit pairs of points with $d\wR/dk=0$. If they do, it means that there are stable modes with vanishing group velocity which, so to speak, remain in the original region. This implies that the corresponding instability is convective, because the modes with vanishing group velocity remain in the system. If there are no pairs of points on the \edit{real-valued} branch with $d\wR/dk=0$, then the modes remaining in the system have turned unstable, so the instability has become absolute. For case G3, the instability has already turned absolute, since there are no points with horizontal tangent in Fig.~\ref{fig:DR_branches_w}. In Appendix~\ref{app:absolute}, we show an explicit example, intermediate between cases G2 and G3, in which the instability is convective.  Generally, for a very weak crossing, the instability is convective, and only becomes absolute as the crossing gets deeper. The physics is again particularly clear in the resonant picture: when the instability develops for very weak crossings, it involves only modes resonant with neutrinos close to the crossing region, with $v\sim +1$. Therefore, these modes tend to escape the system as they grow, implying an initially convective instability.
    
    \item[{\bf 3.}] {\bf Moderate crossing (G4, G5)}. \\
    As the crossing grows deeper, the asymmetry between the flipped region, close to $\uR=+1$, and the unflipped one, on the side of $\uR=-1$, becomes less pronounced. The unstable modes no longer both escape the light cone at $\uR=+1$; they escape on opposite sides. When this happens, the solutions at $k=0$, which we can find explicitly and were real before, become complex.
    
    In case G4, the complex modes in the region $\uR<-1$ do not continue uninterrupted, since they briefly merge with the \edit{real-valued} branch and reappear, in a pair of superluminal branching points.
    In case G5, the pair of branching points has vanished and the entire interval of wavenumbers between the two subluminal critical points has turned unstable. The \edit{real-valued} branch close to $\uR=-1$ remains.
    
    \item[{\bf 4.}] {\bf Deep crossing (G6)}. \\ 
    As the crossing deepens, the asymmetry between the positive and negative $G_v$ regions becomes even less pronounced. \edit{When the phase velocity of the Goldstone mode eventually becomes subluminal, i.e., when $|G_0 / G_1| <1$,} the real-valued branch close to $\uR=-1$ disappears as well, and only complex solutions remain. \edit{The Goldstone mode, which itself is always stable, is now part of a new branch of Landau-damped modes which appears close to $\uR=-1$. These damped solutions will merge smoothly with the unstable modes when $v_c<0$, at which point the slope of the complex branch ($\partial \omega_R/\partial k$) will become negative. This behavior results from backward neutrinos ($v<0$) becoming resonant with the flavor waves as the crossing gets deeper.}
\end{itemize}

\edit{As shown in Table~\ref{tab:ex_params}, the angular distributions G1--G6 fulfill $N_s=0$. This can be deduced from  Fig.~\ref{fig:DR_branches_w}: in each of these cases, there is a superluminal real-valued mode with $k=G_1$, corresponding to the trivial solution of Eq.~\eqref{eq:goldstone}, i.e., $\omega=G_0$. 
The Nyquist criterion thus holds in each of these cases as stated in \cite{Fiorillo:2023hlk}. 

As a counterexample, we briefly turn to an angular distribution for which $\Phi (u)$, as defined in Eq.~\eqref{eq:dr}, has additional zeroes on the real axis. This distribution is shown in the top panel of Fig.~\ref{fig:nyquist_exception}, and belongs to a slightly different family:
 \begin{equation}\label{eq:Nyquist_exception}
     G_v = 0.21 -\frac{0.1}{a}\exp\left( -\frac{(v-1)^2}{2b^2}\right)\,,
 \end{equation}
with $a=0.35$ and $b=0.15$. Thus, we consider an isotropic $\nu_e$ distribution instead of a Gaussian. The corresponding branches of the dispersion relation are plotted in the bottom panel of Fig.~\ref{fig:nyquist_exception}. We see that the real-valued branch of the dispersion relation intersects with the vertical line $k=G_1$ at three different values of $\omega$. Hence, the dispersion relation of the homogeneous mode admits two  real-valued superluminal solutions besides the Goldstone mode described in Sec.~\ref{sec:homogeneous_mode}. This means that $W=1$, $N_s=2$ and, according to our corrected Nyquist criterion, $N=W-N_s/2=0$. Indeed, there are no unstable modes with $k=G_1$.

 \begin{figure}[ht]
 \centering
 \includegraphics[width=8cm]{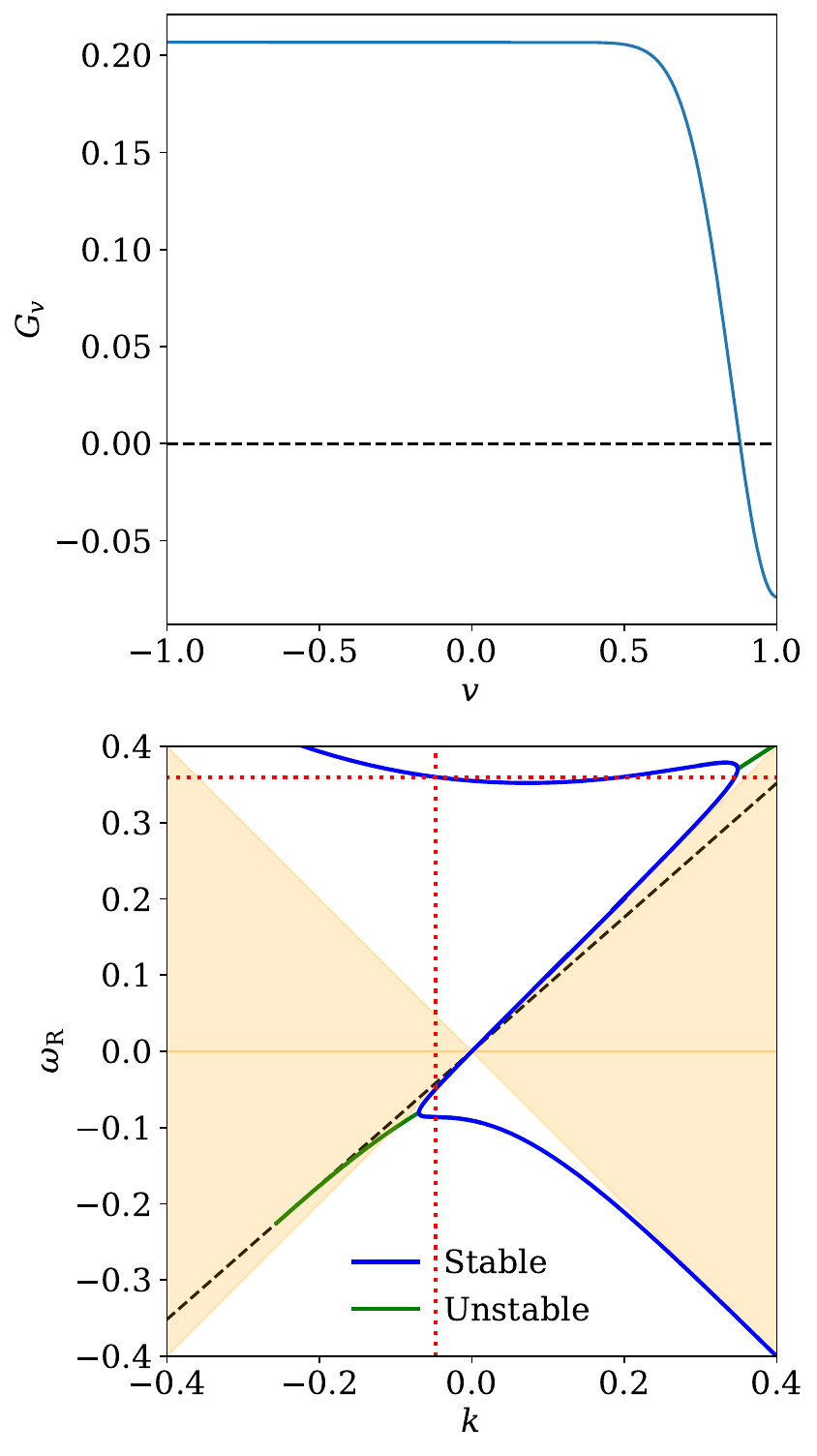}
 \caption{\color{black}\textit{Top:} Example of an angular distribution $G_v$, given in Eq.~\eqref{eq:Nyquist_exception}, for which $N_s\neq 0$. \textit{Bottom:} Branches of the corresponding dispersion relation. The real-valued branch is shown in blue, the unstable one in green. The stable branch intersects with $k=G_1$ (red dotted) at $N_s+1=3$ points (we recall that there is always an additional superluminal Goldstone mode).}
 \label{fig:nyquist_exception}
 \end{figure}}
 
Overall, the evolution can be understood intuitively with the resonance picture previously developed \cite{Fiorillo:2024bzm, Fiorillo:2024uki}: when the crossing appears, waves can resonantly grow close to $\uR=1$, resonantly extracting energy from neutrinos in the flipped region. Case G3 is precisely of this type, with a very weak instability. As the crossing gets deeper, the width of the resonance describing the particle-wave interaction increases, so ever more waves with $\uR\gg 1$ can also grow, until even modes with $\uR<-1$ can nonresonantly grow. At this stage, the distinction between a dominant and a flipped region in $G_v$ becomes superfluous: they contain comparable amounts of lepton number, so there can be growth for modes on both sides of $\uR<\vc$ or $\uR>\vc$.

We prove these properties in Sec.~\ref{sec:General} beyond a single numerical example. One particularly vexing question is if the real-valued superluminal mode could have more than two critical points, equivalent to asking if there could be an unstable mode that both begins and ends outside the light cone on two different superluminal critical points. No such instance arises in any of the cases we have studied, although we have not been able to prove in a formal sense that this cannot happen.

\section{General Properties}
\label{sec:General}

We now turn to a general discussion of the instability properties of single-crossed angular distributions. Many properties that were proved using the Nyquist criterion for the special case of the homogeneous mode ($K=0$) can be understood from a different perspective if one considers modes with any $K$. Without pretense of formal rigor, we will assume some sort of continuity such that for varying $K$, a given branch, real or unstable, does \edit{not} disappear abruptly, with the exception of it ending on a branch cut. By this simple assumption, the properties of the unstable range of wavenumbers follow completely. We use the framework of the causal dispersion relation in the Landau picture that ensures continuity. In the often-used normal-modes picture, excitations with vanishing growth rate can appear and disappear abruptly, somewhat hiding their physical nature. 

We first study the properties of modes in three separate regimes that can be understood with simple arguments. Later, we merge them to the complete picture that we have anticipated in the introduction.

\subsection{Subluminal modes}

In the Landau framework, subluminal collective modes are either Landau damped or unstable. In the normal-modes approach, on the other hand, the subluminal range is populated with Case-van Kampen modes, corresponding to individual neutrino modes slightly perturbed by the interaction. Pairs of such modes can merge to form unstable ones. In our framework, without an angular crossing, only Landau-damped modes exist~\cite{Fiorillo:2024uki}. They can appear abruptly only when their phase velocity is $\mathrm{Re}(u)=\pm 1$, as in the top left panel of Fig.~\ref{fig:DR_branches}, caused by the dispersion relation having a branch cut along these lines, as we have discussed. 

Once $G_v$ develops a crossing, instabilities appear. The transition between unstable and Landau-damped modes must happen where $\mathrm{Im}(u)=0$ for a real $k$. This is a very special point because it must have $|u|<1$ and $\mathrm{Im}(u)=0$, while the wavenumber(s) given by  Eq.~\eqref{eq:k_function_u} must be real. There is only one phase velocity for which this can happen, the crossing velocity $u=\vc$ \cite{Fiorillo:2024uki}. For real $u$, the integrals $J_n$ are purely real and their imaginary parts are proportional to $G_{v=u}$ that vanishes at $u=\vc$. Physically, this special mode resonates with neutrinos at the crossing, which carries zero lepton number. Furthermore, for $u=\vc$, the argument of the square root in Eq.~\eqref{eq:k_function_u} is positive. Therefore, a crossing guarantees two critical points for $u=\vc$.

Moreover, the solutions $k_\pm$, corresponding to the two sign choices in Eq.~\eqref{eq:k_function_u}, actually have opposite signs. This happens if
\begin{equation}
    \left(\frac{J_0+J_2}{2}\right)^2-J_1^2>\left(\frac{J_0-J_2}{2}\right)^2,
\end{equation}
implying that we need
\begin{equation}\label{eq:J-relation}
    \frac{J_2}{J_0}>\left(\frac{J_1}{J_0}\right)^2.
\end{equation}
Since $J_n$ for $u=\vc$ corresponds to integrating $v^n$ over the positive-definite distribution $G_v/(\vc-v)$, the left-hand side is the average $\langle v^2\rangle$, while the right-hand side is $\langle v\rangle^2$. For a positive-definite distribution, $\langle v^2\rangle > \langle v\rangle^2$, so Eq.~\eqref{eq:J-relation} is always verified.

Two separate transitions from Landau-damped to unstable modes on opposite sides of $k=0$ mean that they must occur on two separate Landau-damped branches. Therefore, we learn that the uncrossed $G_v$ must show two such branches, so that once a crossing develops, both of them can develop a transition to instability.

In summary, for an uncrossed $G_v$, two branches of Landau-damped modes exist which begin at $u=-1$ and end at $u=+1$. Once a crossing develops, each of them changes to unstable at $u=\vc$. Importantly, the unstable modes do not end on the light cone, so they can become superluminal, unlike the Landau-damped ones. Hence the unstable branch can escape the subluminal range, at $u=+1$ or $u=-1$; which of them depends on the relative signs of $G_v$ before and after the crossing. Qualitatively, the unstable modes are on the \textit{weak} side of the crossing, i.e., the one that contains less lepton number. In principle, the Nyquist criterion or the explicit expressions for the growth rate~\cite{Fiorillo:2024uki} allows one to predict on which side of the crossing this happens, but we will later provide an explicit criterion based on a different argument.

\subsection{Modes close to the light cone}

We now turn to superluminal modes close to the light cone ($u\simeq \pm 1$) and focus on modes with very large $k$, where the dispersion relation is particularly simple. Starting from Eq.~\eqref{eq:k_function_u}, for $u$ close to $\pm 1$, the term $J_0-J_2$ converges (the numerator vanishes for $v=\pm 1$), whereas the term under the square root diverges. Thus, for $k\to +\infty$, a superluminal mode can only exist if that term is positive and tends to infinity, so that
\edit{\begin{equation}
    {k^2}=\frac{\left(J_0+J_2-2J_1\right)(J_0+J_2+2J_1)}{4}\,. 
\end{equation}}%
For $u=1+\delta$ with $\delta\ll 1$, the first factor converges and equals $(J_0+J_2-2J_1)\to G_0-G_1$, whereas the second factor instead diverges. 

Following the procedure outlined in Ref.~\cite{Fiorillo:2024bzm}, we can extract the leading singularity as
\begin{equation}
    J_0+J_2+2J_1\simeq 4 G(1) \log\left[\frac{2}{\delta}\right],
\end{equation}
where $G(1)$ means $G_v$ evaluated at $v=1$, not to be confused with the first moment $G_1$ (see footnote~1).
So the dispersion relation for modes close to the light cone $u\simeq +1$ at $k\to \infty$ is
\edit{\begin{equation}
    u\simeq 1+2\exp\left[-\frac{k^2}{ G(1)(G_0-G_1)}\right].
\end{equation}}%
To be a consistent solution with $\delta \to 0$ as $k\to \infty$, we must have $G(1) (G_0-G_1)>0$. For an uncrossed $G_v$ this is guaranteed, since $G_v$ is positive and always $G_0-G_1>0$ because $1-v>0$ for any $-1<v<1$. 

We can perform the same operations for $u\simeq -1-\delta$, close to the other side of the light cone, obtaining
\edit{\begin{equation}
    u\simeq -1-2\exp\left[-\frac{k^2}{ G(-1)(G_0+G_1)}\right].
\end{equation}}%
Again, for $G_v$ uncrossed, $G(-1) (G_0+G_1)>0$, so these modes exist. Hence, there are two families of real-valued superluminal modes starting at $k\to -\infty$ and reaching $k\to +\infty$, one with $u>1$, the other with $u<-1$. These are the two branches shown in case G1 in Figs.~\ref{fig:DR_branches} and \ref{fig:DR_branches_w}.

Introducing a $G_v$ crossing makes one of $G(\pm1)$ change sign, while $G_0\pm G_1$ at first remains positive, since the bulk of the distribution is not strongly affected by flipping a small region. Therefore, one of the two superluminal branches disappears, depending on which of $G(\pm1)$ changes sign. Following our examples in Sec.~\ref{sec:Example}, we take $G(1)<0$, so it is the superluminal branch with $u>1$ that disappears. It will be replaced by the unstable subluminal branch that crosses the light cones and becomes superluminal. Once again, \hbox{$u>\vc$} (the weak side of the crossing) is where an unstable mode initially appears.

Finally, if the distribution is so distorted that both $G(1)(G_0-G_1)<0$ and $G(-1)(G_0+G_1)<0$, both of the modes starting at the light cone at $k\to \pm\infty$ disappear, corresponding to G6 in Fig.~\ref{fig:DR_branches}.

\subsection{Superluminal modes}

Superluminal modes can be either stable or unstable, the latter only if there is a crossing. By our principle of continuity, the only possible superluminal modes must be the ones starting at the light cone at $k\to\pm \infty$. When $G_v$ develops a crossing, on one side of the light cone, the one at $u<-1$ for $G(1)(G_0-G_1)<0$, the real-valued superluminal mode survives. On the other side of the crossing are unstable modes originating from within the light cone. These unstable superluminal modes always come with a complex conjugate, so while within the light cone, there is only a single branch, but as soon as they escape, a damped complex conjugate suddenly appears. This behavior does not contradict our principle of continuity because the branch cut at $\mathrm{Re}(u)=\pm 1$ and $\mathrm{Im}(u)<0$ permits modes to appear discontinuously on this line.

The behavior of these unstable modes that ``leak out'' of the light cone crucially depends on the properties of $G_v$. As we have seen, these modes come from opposite sides of $k=0$, a point of infinite phase velocity, that may still be real or complex. At $k=0$, the dispersion relation is completely algebraic of the form
\edit{\begin{equation}
    \left(\frac{G_0}{\omega}-1\right)\left(\frac{ G_2}{\omega}+1\right)-\frac{ G_1^2}{\omega^2}=0
\end{equation}
with solutions
\begin{equation}
    \omega=\frac{1}{2}\left[ (G_0-G_2) \pm \sqrt{(G_0-G_2)^2-4G_1^2}\right].
\end{equation}}%
For $(G_0-G_2)^2\geq 4G_1^2$, this frequency is real so that on opposite sides of $k=0$, the unstable modes emerging from the light cone must turn real. It was first shown in Refs.~\cite{Yi:2019hrp, Fiorillo:2024uki} that this can happen only at superluminal points that must have $\partial \Phi(u,k)/\partial u=0$. \edit{Here, $\Phi(u,k)=0$ is the dispersion relation for arbitrary $k$, such that $\Phi (u,G_1)=0$ yields Eq.~\eqref{eq:dr}. The condition $\partial \Phi(u,k)/\partial u=0$ thus marks the critical points where the unstable branches merge with a superluminal real-valued 
branch}. While these two equations provide two conditions for $u$ and $k$, they involve integral functions so that the critical points usually cannot be found analytically. 

One way to proceed is to graph $k$ as a function of $u$ for $u>1$ using Eq.~\eqref{eq:k_function_u} that does not require solving a transcendental equation. From this plot, one can find the $k$ range where a stable mode exists, and by complementarity, also the range where instead an unstable mode exists. Indeed, Fig.~\ref{fig:DR_branches} shows clearly that the unstable branches in cases G3 and G4 disappear at the points where the \edit{real-valued} branches have vertical tangents. These are precisely the superluminal critical points discussed here.

For a single-crossed $G_v$, these superluminal critical points always come in pairs. Their ways of appearance are exemplified by cases G3 and G4. In the former, the two unstable modes on opposite sides of $k=0$ must disappear at two critical points with $\uR>1$ on the other sides of $k=0$. In case G4, the unstable modes for $\uR<-1$ exhibit a pair of critical points on the same \edit{real-valued} branch, which has a characteristic S shape. One may wonder whether multiple such pairs might exist. We have not been able to find a corresponding $G_v$, but also not a proof of nonexistence.

For $(G_0-G_2)^2<4G_1^2$, the frequency at $k=0$ turns complex, marking the transition between cases G3 and G4, when the unstable branches cross all the way through $k=0$. Instead, throughout cases G5 and G6, the entire $k$ interval between the two subluminal critical points has turned unstable, not only for $k=0$. This is the situation in which the original Nyquist criterion of Eq.~\eqref{eq:OldNyquist} is both necessary and sufficient. However, we could not find an analytic condition to mark the transition between G4 and G5, because again the superluminal critical points cannot be easily identified. As described earlier, they can instead be found by plotting the \edit{real-valued} branches and checking if there are points with $d\uR/dk\to +\infty$.

\subsection{Summary}

To summarize, we have found that for an uncrossed $G_v$, there are two branches of Landau-damped modes beginning and ending on the light cone, and two superluminal real-valued ones, also beginning and ending there. 

For a weak crossing, one of the real-valued superluminal modes vanishes, the one at $u>1$ if $G(1)(G_0-G_1)<0$, or the one at $u<-1$ if $G(-1)(G_0+G_1)<0$. At the same time, the Landau-damped modes become unstable and escape the light cone, where they return to stable at two critical superluminal points on opposite sides of $k=0$. So there are two intervals of instability $(k_1,k_2)$ and $(k_3,k_4)$, as anticipated in the introduction. The positions $k_1$ and $k_4$, the subluminal critical points, are given by $k_\pm$ from Eq.~\eqref{eq:k_function_u} for $u=\vc$. On the other hand, the positions $k_2$ and $k_3$, the superluminal critical points, cannot be identified by an explicit expression, but can be found by plotting the function Eq.~\eqref{eq:k_function_u} for $u>1$ (or $u<-1$).

For a deep $G_v$ crossing, defined by the condition $(G_0-G_2)^2<4G_1^2$, the unstable branches pass through the point $k=0$ with unstable modes with both positive and negative $\uR$. In particular, the entire subluminal range $-1<\uR<1$ now contains unstable modes. As the crossing gets even deeper, the critical superluminal points disappear and the modes remain unstable throughout the entire wavenumber interval.

Finally, when the distribution is distorted so much that even the second among the condition $G(1)(G_0-G_1)<0$ and $G(-1)(G_0+G_1)<0$ is verified, the real-valued mode on the other side of the light cone disappears entirely.

Our conclusions qualitatively agree with the ones of Ref.~\cite{Yi:2019hrp}. In addition, we provide explicit inequalities that reveal which regime follows from a given $G_v$. Most importantly, we connect these results with the recently developed plasma-physics language and explain how the results from the Nyquist criterion, using entirely different methods from complex analysis, match completely the results from our more general treatment here. Further, we show how the unstable modes' first appearance is essentially ``inherited'' from the Landau-damped modes that exist even before the formation of a crossing. These physical excitations do not appear in the more traditional normal-modes approach.

\section{Summary and Conclusions}
\label{sec:Conclusion}

The question that has motivated this study was to identify the conditions for the homogeneous mode of an infinitely homogeneous neutrino gas to turn unstable. A first answer provided by two of us~\cite{Fiorillo:2023hlk}, using a Nyquist criterion for its dispersion relation, was actually incomplete as we have here clarified. We have stressed that the homogeneous mode is inseparably connected with the properties of the general dispersion relation and plays no special role. Its status as a prototype of flavor conversion mainly derives from the simplicity of its nonlinear evolution, a simplicity that however is deceiving because it descends from the many symmetries forced upon it. Therefore, we emphasize that the homogeneous mode can only be understood in the context of the dispersion relation for modes with arbitrary wavenumbers.

The formal properties of the dispersion relation were studied before~\cite{Yi:2019hrp, Capozzi:2019lso}. Our aim was to connect them with the plasma-physics inspired picture that we have recently developed \cite{Fiorillo:2023mze, Fiorillo:2023hlk, Fiorillo:2024qbl, Fiorillo:2024bzm, Fiorillo:2024uki}. For a single-crossed angular distribution, the instability develops close to that side of the light cone where the angular crossing first appears, $v=1$ in our examples. We have connected this finding with the resonant picture of the instability: it involves modes resonant with the neutrinos close to the crossing region, and therefore involves initially only a few modes outside of the light cone. These unstable modes are directly related to the Landau-damped ones below the light cone, which appear even without a crossing, but a nearly vanishing $G_v$ at $v=1$. They could not be found in previous works that used the dispersion relation for normal modes, not physical ones \cite{Fiorillo:2024bzm, Fiorillo:2024uki}, the formal difference being Landau's $i\epsilon$ prescription in integrals of the type of Eq.~\eqref{eq:landau_dispersion_integrals}. 

Thus the Landau-damped modes, which appear even for the uncrossed distribution, are the progenitors of the fast unstable modes. Intriguingly, the stable branches of real eigenmodes, which also appear for an uncrossed distribution, have played mainly a spectator role in this work, without directly influencing the fast unstable modes. However, as recently shown \cite{Fiorillo:2024pns}, these stable modes are instead the progenitors of the slow unstable modes, once a small vacuum energy splitting between neutrinos and antineutrinos is introduced.

Following earlier work \cite{Yi:2019hrp}, we used a generic example for which the depth of the angular crossing is controlled by a single parameter $a$. For increasing $a$, the crossing grows deeper, and the interval of unstable wavenumbers grows, as well as that of unstable phase velocities, until it encompasses $k=0$ and therefore infinite phase velocity. We provide explicit conditions for this transition. Finally, the depth grows to a point where the two intervals of unstable wavenumbers merge to a single one. While we could not identify a simple criterion for this to happen, and whether the homogeneous mode $k=G_1$ falls into such intervals, it follows graphically from the existence and shape of the \edit{real-valued} branch $\omega(k)$, i.e., the number of critical points with vertical tangent.

The Nyquist criterion \cite{Fiorillo:2023hlk} shown in Eq.~\eqref{eq:OldNyquist} remains valid as a nontrivial necessary condition. But it does not encompass the possibility that pairs of superluminal critical points appear, creating an ``interval of stability'' in which the homogeneous mode may fall and then become stable despite the Nyquist criterion. This possibility is best understood by looking directly at the global properties of the modes and their phase velocity. \edit{We have showcased this scenario in Sec.~\ref{sec:Example} (Fig.~\ref{fig:nyquist_exception}), and tested that the relation $N=W-N_s/2$ holds for all of our angular distributions.} 

Our discussion was mostly based on formal properties of the dispersion relation and its connection with the resonant picture of the instability\edit{. We have illustrated how unstable modes emerge from Landau-damped ones and provided a method to count the number of complex-$\omega$ solutions associated with a given $k$. However, our treatment of the dispersion relation is not exhaustive. More work is needed to analytically describe some of the behavior  that we observe: we have not been able to find a condition for the merger of the two unstable branches of the dispersion relation, i.e., the transition from our case G4 to G5. Further research could also establish whether the superluminal real-valued branches may have more than two critical points.} 

\edit{Our examples do show that, for very shallow angular crossings, } only two narrow intervals of large $k$ are actually unstable. Using the parameter $\epsilon=(n_\nu-n_{\overline{\nu}})/(n_\nu+n_{\overline{\nu}})$, where $n_\nu$ and $n_{\overline{\nu}}$ are the neutrino and antineutrino densities \cite{Fiorillo:2024pns}, unstable modes have $k\sim \epsilon$. \edit{This could suggest that flavor instabilities are initially driven by inhomogeneous modes with low associated growth rates \cite{Fiorillo:2024wej, Fiorillo:2024bzm, Fiorillo:2024uki}, similar to case~G3. Our cases G4--G6 may be less relevant for practical astrophysical environments. A self-consistently evolving system likely cannot develop a strong instability, which on the contrary is a signature of inconsistency \cite{Johns:2023jjt, Johns:2024dbe, Fiorillo:2024wej}. 
}
\edit{Within this picture, the homogeneous mode is phenomenologically secondary: for small angular crossings, the intervals of unstable $k$ are very narrow and likely exclude $k=G_1$. However, the homogeneous mode retains its theoretical interest as an exact nonlinear solution of the equations of motion. In this sense, its place in flavor conversion research is comparable to that of solitons in plasma physics, as a completely integrable solution~\cite{Johns:2019izj, Padilla-Gay:2021haz, Fiorillo:2023mze, Fiorillo:2023hlk} with nontrivial collective features, e.g.~the conservation of the neutrino-neutrino interaction energy~\cite{Fiorillo:2024fnl}. 
}

\section*{Acknowledgments}

DFGF is supported by the Alexander von Humboldt Foundation (Germany). MGG acknowledges support from the European Union (ERC, ANET, Project No. 101087058, PI: I. Tamborra). Views and opinions expressed are those of the authors only and do not necessarily reflect those of the European Union or the European Research Council. Neither the European Union nor the granting authority can be held responsible for them. GGR acknowledges partial support by the German Research Foundation (DFG) through the Collaborative Research Centre ``Neutrinos and Dark Matter in Astro- and Particle Physics (NDM),'' Grant SFB-1258-283604770, and under Germany’s Excellence Strategy through the Cluster of Excellence ORIGINS EXC-2094-390783311.

\appendix

\section{Absolute and convective instabilities}
\label{app:absolute}

The distinction between absolute and convective instabilities in the context of fast flavor waves was first made in Ref.~\cite{Capozzi:2017gqd} and further developed in Ref.~\cite{Yi:2019hrp}. While there is not much to add on the formal level, here we connect this general distinction with the resonant picture of the fast instability, showing that this notion provides intuition as to why the instability is convective for a weak crossing and only for a sufficiently deep one becomes absolute. To discuss these issues, we first need to review the concepts of convective vs.\ absolute and their physical meaning.

Formally, an instability is absolute if a localized perturbation spawns exponential temporal growth everywhere. This is assured for a plane wave, which by definition is completely delocalized, if $\omega(k)$ is unstable. On the other hand, all unstable modes can have group velocities so large along some direction that the unstable components escape from the localized region of the initial perturbation, leading to growth only somewhere else. In this case, the instability is said to be convective. This distinction may have been first pointed out by Landau and Lifshitz in the context of hydrodynamical instabilities~\cite{landau1987fluid}, but it was Sturrock~\cite{sturrock1958kinematics} who studied this topic in detail in the context of kinetic instabilities, i.e., caused by particles streaming with different velocities. 

Evidently, the distinction between absolute and convective instabilities must depend on the group velocities of the unstable waves. Here we do not try to justify Sturrock's results, but merely affirm their plausibility. A general criterion for an instability to be absolute is that $\mathrm{Im}(\omega)>0$ with $d\omega/dk=0$. Such a mode has vanishing group velocity, and therefore relaxes locally, so this condition is not too surprising, although its formal justification is not immediate~\cite{sturrock1958kinematics}.

\begin{figure}[ht]
\centering
\includegraphics[width=0.9\columnwidth]{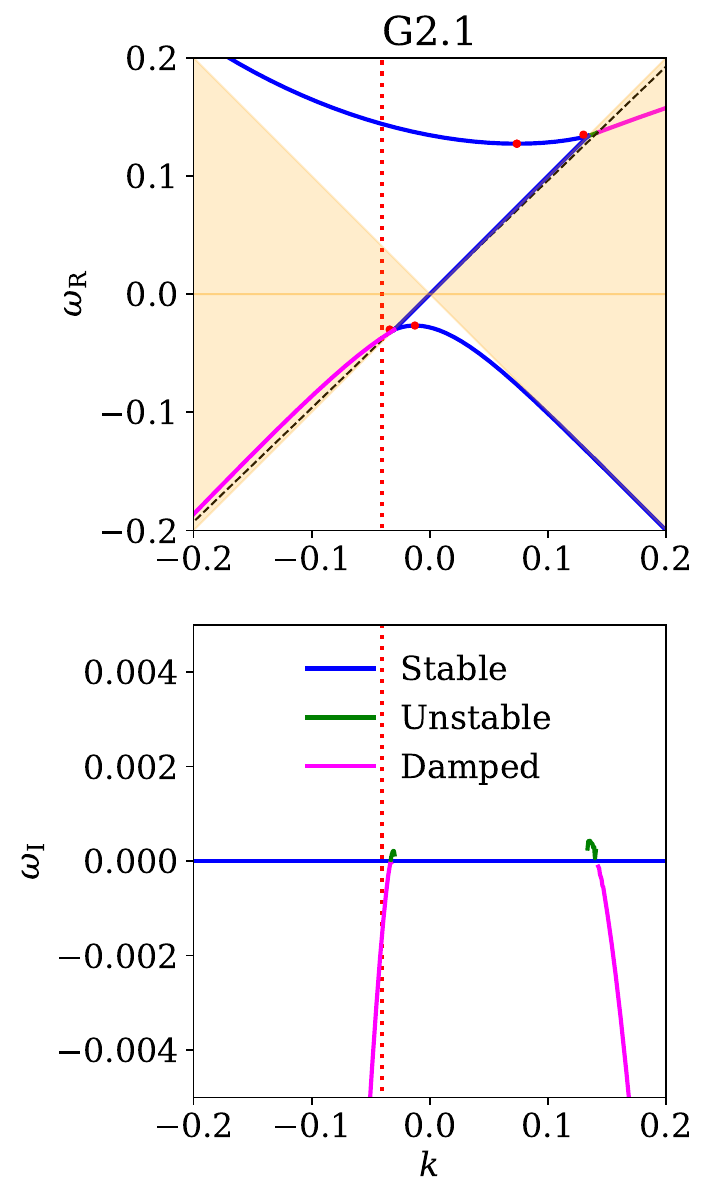}
\caption{Branches of the dispersion relation for a $G_v$ of the form of Eq.~\eqref{eq:ang_spectrum} with $a=0.856$. The \edit{real-valued} branches (blue) have 
four critical points with $d\wR/dk =0$ (highlighted in red), implying that the instability is convective.\label{fig:tangent}}
\end{figure}

In the context of fast flavor conversions, such as our numerical example
in Sec.~\ref{sec:Example}, when in a family of angular distributions, these first develop a weak crossing, the instability is always convective. This property can be understood by considering how the structure of the modes changes when an angular crossing first appears. As seen in case G2 of Fig.~\ref{fig:DR_branches_w}, when $G_v$ vanishes at $v=1$ and the crossing is about to appear, both \edit{real-valued} branches have a point of horizontal tangent ($d\omega/dk=0$). By continuity, when the crossing first develops, these points cannot disappear, and the two previously disconnected \edit{real-valued} branches merge to a single real one. One easily sees graphically that when this happens, two additional points with horizontal tangent must appear. We illustrate this circumstance in Fig.~\ref{fig:tangent}, where we choose an angular distribution ``G2.1'' intermediate between cases G2 and G3 of the main text. We highlight the four points with horizontal tangent, corresponding to two pairs of stable modes with vanishing group velocity. So at this stage the modes with $d\omega/dk=0$ are stable; the instability is therefore of a convective nature. 

Only as the crossing grows deeper, each pair of solutions with $\omega$ real and $d\omega/dk=0$ disappears when the two points with horizontal tangent belonging to each pair merge. From analysis, we know that this happens when $d\omega/dk=d^2\omega/dk^2=0$ simultaneously. When the crossing is made even deeper, the two stable modes with zero group velocity have disappeared and moved to the complex plane, which means that one solution with $\mathrm{Im}(\omega)>0$ and $d\omega/dk=0$ has now appeared. The instability has therefore turned absolute. This transition was first noted in Ref.~\cite{Yi:2019hrp}, which also showed that the points with $d\omega/dk=0$, which are present when the instability is convective, serve as onset points for branches of complex $k$ for real $\omega$, in the same way as the superluminal critical points with $dk/d\omega=0$ correspond to the appearance of branches of complex $\omega$ for real $k$.

A simple interpretation can be given as to the convective nature of the instability for our single-crossed spectra. When the instability first appears, it is caused by the resonant emission of flavor waves from the neutrinos belonging to the flipped region~\cite{Fiorillo:2024bzm, Fiorillo:2024uki}. This process bears strong resemblance to beam-plasma instabilities, when a beam of fast particles is passed through a background plasma. The instability then ensues when the particles in the beam resonantly emit plasma waves to get rid of their excess energy, until the velocity distribution flattens. In the fast flavor case, when the angular crossing first develops, the angular distribution can be regarded as a beam of negative lepton number passing through a medium with positive lepton number. In this case a key role is played by lepton-number conservation~\cite{Johns:2024bob, Fiorillo:2024bzm}, which would impede the conversion without the beam. Instead, as discussed in Refs.~\cite{Fiorillo:2024bzm, Fiorillo:2024uki}, the beam of negative lepton number gets rid of it by resonant emission of flavor waves, ultimately erasing the angular crossing, an effect observed numerically~\cite{Zaizen:2022cik, Nagakura:2022kic} and predicted for a simple setup by quasi-linear theory \cite{Fiorillo:2024qbl}. 

In this picture, when the beam of lepton number (or, in the plasma analogy, the beam of energetic particles) first appears, it resonantly produces waves that move with the same phase velocity as the particles, away from the region in which they were formed. While the phase velocity does not directly determine the motion of wavepackets, it is intuitively clear that the growing waves are initially leaving the system, and the instability is therefore convective. For deeper crossings, the resonant growth ultimately encompasses also modes that do not move with a sufficiently large group velocity. We do not go here into the question of what are the physical implications of the instability being absolute vs.\ convective.

\bibliographystyle{JHEP}
\bibliography{References}

\providecommand{\href}[2]{#2}\begingroup\raggedright\begin{thebibliography}{10}

\bibitem{Morinaga:2021vmc}
T.~Morinaga, \emph{{Fast neutrino flavor instability and neutrino flavor lepton number crossings}}, \href{https://doi.org/10.1103/PhysRevD.105.L101301}{\emph{Phys. Rev. D} {\bfseries 105} (2022) L101301} [\href{https://arxiv.org/abs/2103.15267}{{\ttfamily 2103.15267}}].

\bibitem{Dasgupta:2021gfs}
B.~Dasgupta, \emph{{Collective Neutrino Flavor Instability Requires a Crossing}}, \href{https://doi.org/10.1103/PhysRevLett.128.081102}{\emph{Phys. Rev. Lett.} {\bfseries 128} (2022) 081102} [\href{https://arxiv.org/abs/2110.00192}{{\ttfamily 2110.00192}}].

\bibitem{Johns:2024bob}
L.~Johns, \emph{{Ergodicity demystifies fast neutrino flavor instability}},  \href{https://arxiv.org/abs/2402.08896}{{\ttfamily 2402.08896}}.

\bibitem{Fiorillo:2024bzm}
D.F.G.~Fiorillo and G.G.~Raffelt, \emph{{Theory of neutrino fast flavor evolution. Part~I. Linear response theory and stability conditions.}}, \href{https://doi.org/10.1007/JHEP08(2024)225}{\emph{JHEP} {\bfseries 08} (2024) 225} [\href{https://arxiv.org/abs/2406.06708}{{\ttfamily 2406.06708}}].

\bibitem{Fiorillo:2024uki}
D.F.G.~Fiorillo and G.G.~Raffelt, \emph{{Theory of neutrino fast flavor conversions. Part II. Solutions at the edge of instability}},  \href{https://arxiv.org/abs/2409.17232}{{\ttfamily 2409.17232}}.

\bibitem{Capozzi:2019lso}
F.~Capozzi, G.~Raffelt and T.~Stirner, \emph{{Fast Neutrino Flavor Conversion: Collective Motion vs. Decoherence}}, \href{https://doi.org/10.1088/1475-7516/2019/09/002}{\emph{JCAP} {\bfseries 09} (2019) 002} [\href{https://arxiv.org/abs/1906.08794}{{\ttfamily 1906.08794}}].

\bibitem{Johns:2019izj}
L.~Johns, H.~Nagakura, G.M.~Fuller and A.~Burrows, \emph{{Neutrino oscillations in supernovae: angular moments and fast instabilities}}, \href{https://doi.org/10.1103/PhysRevD.101.043009}{\emph{Phys. Rev. D} {\bfseries 101} (2020) 043009} [\href{https://arxiv.org/abs/1910.05682}{{\ttfamily 1910.05682}}].

\bibitem{Padilla-Gay:2021haz}
I.~Padilla-Gay, I.~Tamborra and G.G.~Raffelt, \emph{{Neutrino Flavor Pendulum Reloaded: The Case of Fast Pairwise Conversion}}, \href{https://doi.org/10.1103/PhysRevLett.128.121102}{\emph{Phys. Rev. Lett.} {\bfseries 128} (2022) 121102} [\href{https://arxiv.org/abs/2109.14627}{{\ttfamily 2109.14627}}].

\bibitem{Fiorillo:2023mze}
D.F.G.~Fiorillo and G.G.~Raffelt, \emph{{Slow and fast collective neutrino oscillations: Invariants and reciprocity}}, \href{https://doi.org/10.1103/PhysRevD.107.043024}{\emph{Phys. Rev. D} {\bfseries 107} (2023) 043024} [\href{https://arxiv.org/abs/2301.09650}{{\ttfamily 2301.09650}}].

\bibitem{Fiorillo:2023hlk}
D.F.G.~Fiorillo and G.G.~Raffelt, \emph{{Flavor solitons in dense neutrino gases}}, \href{https://doi.org/10.1103/PhysRevD.107.123024}{\emph{Phys. Rev. D} {\bfseries 107} (2023) 123024} [\href{https://arxiv.org/abs/2303.12143}{{\ttfamily 2303.12143}}].

\bibitem{Pehlivan:2011hp}
Y.~Pehlivan, A.B.~Balantekin, T.~Kajino and T.~Yoshida, \emph{{Invariants of collective neutrino oscillations}}, \href{https://doi.org/10.1103/PhysRevD.84.065008}{\emph{Phys. Rev. D} {\bfseries 84} (2011) 065008} [\href{https://arxiv.org/abs/1105.1182}{{\ttfamily 1105.1182}}].

\bibitem{Raffelt:2011yb}
G.G.~Raffelt, \emph{{N-mode coherence in collective neutrino oscillations}}, \href{https://doi.org/10.1103/PhysRevD.83.105022}{\emph{Phys. Rev. D} {\bfseries 83} (2011) 105022} [\href{https://arxiv.org/abs/1103.2891}{{\ttfamily 1103.2891}}]. Erratum: \href{https://doi.org/10.1103/PhysRevD.104.089902}{{\em Phys. Rev. D} {\bf 104} (2021) 089902}.

\bibitem{Mangano:2014zda}
G.~Mangano, A.~Mirizzi and N.~Saviano, \emph{{Damping the neutrino flavor pendulum by breaking homogeneity}}, \href{https://doi.org/10.1103/PhysRevD.89.073017}{\emph{Phys. Rev. D} {\bfseries 89} (2014) 073017} [\href{https://arxiv.org/abs/1403.1892}{{\ttfamily 1403.1892}}].

\bibitem{Padilla-Gay:2022wck}
I.~Padilla-Gay, I.~Tamborra and G.G.~Raffelt, \emph{{Neutrino fast flavor pendulum. II. Collisional damping}}, \href{https://doi.org/10.1103/PhysRevD.106.103031}{\emph{Phys. Rev. D} {\bfseries 106} (2022) 103031} [\href{https://arxiv.org/abs/2209.11235}{{\ttfamily 2209.11235}}].

\bibitem{Fiorillo:2023ajs}
D.F.G.~Fiorillo, I.~Padilla-Gay and G.G.~Raffelt, \emph{{Collisions and collective flavor conversion: Integrating out the fast dynamics}}, \href{https://doi.org/10.1103/PhysRevD.109.063021}{\emph{Phys. Rev. D} {\bfseries 109} (2024) 063021} [\href{https://arxiv.org/abs/2312.07612}{{\ttfamily 2312.07612}}].

\bibitem{Yi:2019hrp}
C.~Yi, L.~Ma, J.D.~Martin and H.~Duan, \emph{{Dispersion relation of the fast neutrino oscillation wave}}, \href{https://doi.org/10.1103/PhysRevD.99.063005}{\emph{Phys. Rev. D} {\bfseries 99} (2019) 063005} [\href{https://arxiv.org/abs/1901.01546}{{\ttfamily 1901.01546}}].

\bibitem{Airen:2018nvp}
S.~Airen, F.~Capozzi, S.~Chakraborty, B.~Dasgupta, G.~Raffelt and T.~Stirner, \emph{{Normal-mode Analysis for Collective Neutrino Oscillations}}, \href{https://doi.org/10.1088/1475-7516/2018/12/019}{\emph{JCAP} {\bfseries 12} (2018) 019} [\href{https://arxiv.org/abs/1809.09137}{{\ttfamily 1809.09137}}].

\bibitem{Dolgov:1980cq}
A.D.~Dolgov, \emph{{Neutrinos in the early universe}}, {\emph{Sov. J. Nucl. Phys.} {\bfseries 33} (1981) 700}. [{\em Yad.\ Fiz.} {\bf 33} (1981) 1309].

\bibitem{Rudsky}
M.A.~{Rudzsky}, \emph{{Kinetic equations for neutrino spin- and type-oscillations in a medium}}, \href{https://doi.org/10.1007/BF00653658}{\emph{Astrophys. Space Sci} {\bfseries 165} (1990) 65}.

\bibitem{Sigl:1993ctk}
G.~Sigl and G.~Raffelt, \emph{{General kinetic description of relativistic mixed neutrinos}}, \href{https://doi.org/10.1016/0550-3213(93)90175-O}{\emph{Nucl. Phys. B} {\bfseries 406} (1993) 423}.

\bibitem{Sirera:1998ia}
M.~Sirera and A.~Perez, \emph{{Relativistic Wigner function approach to neutrino propagation in matter}}, \href{https://doi.org/10.1103/PhysRevD.59.125011}{\emph{Phys. Rev. D} {\bfseries 59} (1999) 125011} [\href{https://arxiv.org/abs/hep-ph/9810347}{{\ttfamily hep-ph/9810347}}].

\bibitem{Yamada:2000za}
S.~Yamada, \emph{{Boltzmann equations for neutrinos with flavor mixings}}, \href{https://doi.org/10.1103/PhysRevD.62.093026}{\emph{Phys. Rev. D} {\bfseries 62} (2000) 093026} [\href{https://arxiv.org/abs/astro-ph/0002502}{{\ttfamily astro-ph/0002502}}].

\bibitem{Vlasenko:2013fja}
A.~Vlasenko, G.M.~Fuller and V.~Cirigliano, \emph{{Neutrino Quantum Kinetics}}, \href{https://doi.org/10.1103/PhysRevD.89.105004}{\emph{Phys. Rev. D} {\bfseries 89} (2014) 105004} [\href{https://arxiv.org/abs/1309.2628}{{\ttfamily 1309.2628}}].

\bibitem{Volpe:2013uxl}
C.~Volpe, D.~V\"a\"an\"anen and C.~Espinoza, \emph{{Extended evolution equations for neutrino propagation in astrophysical and cosmological environments}}, \href{https://doi.org/10.1103/PhysRevD.87.113010}{\emph{Phys. Rev. D} {\bfseries 87} (2013) 113010} [\href{https://arxiv.org/abs/1302.2374}{{\ttfamily 1302.2374}}].

\bibitem{Serreau:2014cfa}
J.~Serreau and C.~Volpe, \emph{{Neutrino-antineutrino correlations in dense anisotropic media}}, \href{https://doi.org/10.1103/PhysRevD.90.125040}{\emph{Phys. Rev. D} {\bfseries 90} (2014) 125040} [\href{https://arxiv.org/abs/1409.3591}{{\ttfamily 1409.3591}}].

\bibitem{Kartavtsev:2015eva}
A.~Kartavtsev, G.~Raffelt and H.~Vogel, \emph{{Neutrino propagation in media: Flavor, helicity, and pair correlations}}, \href{https://doi.org/10.1103/PhysRevD.91.125020}{\emph{Phys. Rev. D} {\bfseries 91} (2015) 125020} [\href{https://arxiv.org/abs/1504.03230}{{\ttfamily 1504.03230}}].

\bibitem{Fiorillo:2024fnl}
D.F.G.~Fiorillo, G.G.~Raffelt and G.~Sigl, \emph{{Inhomogeneous Kinetic Equation for Mixed Neutrinos: Tracing the Missing Energy}}, \href{https://doi.org/10.1103/PhysRevLett.133.021002}{\emph{Phys. Rev. Lett.} {\bfseries 133} (2024) 021002} [\href{https://arxiv.org/abs/2401.05278}{{\ttfamily 2401.05278}}].

\bibitem{Fiorillo:2024wej}
D.F.G.~Fiorillo, G.G.~Raffelt and G.~Sigl, \emph{{Collective neutrino-antineutrino oscillations in dense neutrino environments?}}, \href{https://doi.org/10.1103/PhysRevD.109.043031}{\emph{Phys. Rev. D} {\bfseries 109} (2024) 043031} [\href{https://arxiv.org/abs/2401.02478}{{\ttfamily 2401.02478}}].

\bibitem{Johns:2021qby}
L.~Johns, \emph{{Collisional Flavor Instabilities of Supernova Neutrinos}}, \href{https://doi.org/10.1103/PhysRevLett.130.191001}{\emph{Phys. Rev. Lett.} {\bfseries 130} (2023) 191001} [\href{https://arxiv.org/abs/2104.11369}{{\ttfamily 2104.11369}}].

\bibitem{Xiong:2022zqz}
Z.~Xiong, L.~Johns, M.-R.~Wu and H.~Duan, \emph{{Collisional flavor instability in dense neutrino gases}}, \href{https://doi.org/10.1103/PhysRevD.108.083002}{\emph{Phys. Rev. D} {\bfseries 108} (2023) 083002} [\href{https://arxiv.org/abs/2212.03750}{{\ttfamily 2212.03750}}].

\bibitem{Liu:2023pjw}
J.~Liu, M.~Zaizen and S.~Yamada, \emph{{Systematic study of the resonancelike structure in the collisional flavor instability of neutrinos}}, \href{https://doi.org/10.1103/PhysRevD.107.123011}{\emph{Phys. Rev. D} {\bfseries 107} (2023) 123011} [\href{https://arxiv.org/abs/2302.06263}{{\ttfamily 2302.06263}}].

\bibitem{Lin:2022dek}
Y.-C.~Lin and H.~Duan, \emph{{Collision-induced flavor instability in dense neutrino gases with energy-dependent scattering}}, \href{https://doi.org/10.1103/PhysRevD.107.083034}{\emph{Phys. Rev. D} {\bfseries 107} (2023) 083034} [\href{https://arxiv.org/abs/2210.09218}{{\ttfamily 2210.09218}}].

\bibitem{Johns:2022yqy}
L.~Johns and Z.~Xiong, \emph{{Collisional instabilities of neutrinos and their interplay with fast flavor conversion in compact objects}}, \href{https://doi.org/10.1103/PhysRevD.106.103029}{\emph{Phys. Rev. D} {\bfseries 106} (2022) 103029} [\href{https://arxiv.org/abs/2208.11059}{{\ttfamily 2208.11059}}].

\bibitem{VanKampen:1955wh}
N.G.~Van~Kampen, \emph{{On the theory of stationary waves in plasmas}}, \href{https://doi.org/10.1016/S0031-8914(55)93068-8}{\emph{Physica} {\bfseries 21} (1955) 949}.

\bibitem{Case:1959}
K.M.~{Case}, \emph{{Plasma oscillations}}, \href{https://doi.org/10.1016/0003-4916(59)90029-6}{\emph{Annals of Physics} {\bfseries 7} (1959) 349}.

\bibitem{Sagan:1993es}
D.~Sagan, \emph{{On the physics of Landau damping}}, \href{https://doi.org/10.1119/1.17547}{\emph{Am. J. Phys.} {\bfseries 62} (1994) 450}.

\bibitem{Izaguirre:2016gsx}
I.~Izaguirre, G.~Raffelt and I.~Tamborra, \emph{{Fast Pairwise Conversion of Supernova Neutrinos: A Dispersion-Relation Approach}}, \href{https://doi.org/10.1103/PhysRevLett.118.021101}{\emph{Phys. Rev. Lett.} {\bfseries 118} (2017) 021101} [\href{https://arxiv.org/abs/1610.01612}{{\ttfamily 1610.01612}}].

\bibitem{Landau:1946jc}
L.D.~Landau, \emph{{On the vibrations of the electronic plasma}}, {\emph{J. Phys. (USSR)} {\bfseries 10} (1946) 25}.

\bibitem{thorne2017modern}
K.S.~Thorne and R.D.~Blandford, \emph{Modern classical physics: optics, fluids, plasmas, elasticity, relativity, and statistical physics}, Princeton University Press (2017).

\bibitem{Penrose:1960}
O.~{Penrose}, \emph{{Electrostatic Instabilities of a Uniform Non-Maxwellian Plasma}}, \href{https://doi.org/10.1063/1.1706024}{\emph{Physics of Fluids} {\bfseries 3} (1960) 258}.

\bibitem{schKT}
A.A.~{Schekochihin}, \emph{{Lectures on Kinetic Theory and Magnetohydrodynamics of Plasmas}}, Lecture Notes for the Oxford MMathPhys/MScMTP programme; URL: \url{http://www-thphys.physics.ox.ac.uk/people/AlexanderSchekochihin/KT/2015/KTLectureNotes.pdf} (2024).

\bibitem{Fiorillo:2024qbl}
D.F.G.~Fiorillo and G.G.~Raffelt, \emph{{Fast Flavor Conversions at the Edge of Instability in a Two-Beam Model}}, \href{https://doi.org/10.1103/PhysRevLett.133.221004}{\emph{Phys. Rev. Lett.} {\bfseries 133} (2024) 221004} [\href{https://arxiv.org/abs/2403.12189}{{\ttfamily 2403.12189}}].

\bibitem{Fiorillo:2024pns}
D.F.G.~Fiorillo and G.G.~Raffelt, \emph{{Theory of neutrino slow flavor evolution. Homogeneous medium}},  \href{https://arxiv.org/abs/2412.02747}{{\ttfamily 2412.02747}}.

\bibitem{Johns:2023jjt}
L.~Johns, \emph{{Thermodynamics of oscillating neutrinos}},  \href{https://arxiv.org/abs/2306.14982}{{\ttfamily 2306.14982}}.

\bibitem{Johns:2024dbe}
L.~Johns, \emph{{Subgrid modeling of neutrino oscillations in astrophysics}},  \href{https://arxiv.org/abs/2401.15247}{{\ttfamily 2401.15247}}.

\bibitem{Capozzi:2017gqd}
F.~Capozzi, B.~Dasgupta, E.~Lisi, A.~Marrone and A.~Mirizzi, \emph{{Fast flavor conversions of supernova neutrinos: Classifying instabilities via dispersion relations}}, \href{https://doi.org/10.1103/PhysRevD.96.043016}{\emph{Phys. Rev. D} {\bfseries 96} (2017) 043016} [\href{https://arxiv.org/abs/1706.03360}{{\ttfamily 1706.03360}}].

\bibitem{landau1987fluid}
L.D.~Landau and E.M.~Lifshitz, \emph{Fluid Mechanics: Volume 6}, vol.~6, Elsevier (1987).

\bibitem{sturrock1958kinematics}
P.A.~Sturrock, \emph{Kinematics of growing waves}, {\emph{Physical Review} {\bfseries 112} (1958) 1488}.

\bibitem{Zaizen:2022cik}
M.~Zaizen and H.~Nagakura, \emph{{Simple method for determining asymptotic states of fast neutrino-flavor conversion}}, \href{https://doi.org/10.1103/PhysRevD.107.103022}{\emph{Phys. Rev. D} {\bfseries 107} (2023) 103022} [\href{https://arxiv.org/abs/2211.09343}{{\ttfamily 2211.09343}}].

\bibitem{Nagakura:2022kic}
H.~Nagakura and M.~Zaizen, \emph{{Time-Dependent and Quasisteady Features of Fast Neutrino-Flavor Conversion}}, \href{https://doi.org/10.1103/PhysRevLett.129.261101}{\emph{Phys. Rev. Lett.} {\bfseries 129} (2022) 261101} [\href{https://arxiv.org/abs/2206.04097}{{\ttfamily 2206.04097}}].

\end{thebibliography}\endgroup

\onecolumngrid

\end{document}